\documentclass[%
 reprint,
superscriptaddress,
 hidelinks,
 amsmath,amssymb,
 aps,
]{revtex4-1}

\usepackage{graphicx}
\usepackage{dcolumn}
\usepackage{bm}
\usepackage{xcolor}
\usepackage{hyperref}

\hypersetup{
    colorlinks,
    linkcolor={blue!50!black},
    citecolor={blue!50!black},
    urlcolor={blue!80!black}
}
\begin{document}

\title{Transport Studies in a Gate-Tunable Three-Terminal Josephson Junction}

\author{Gino V. Graziano}
\affiliation{School of Physics and Astronomy, University of Minnesota}
\author{Joon Sue Lee}
\affiliation{California NanoSystems Institute, University of California, Santa Barbara}
\author{Mihir Pendharkar}
\affiliation{Electrical and Computer Engineering Department, University of California, Santa Barbara}
\author{Chris Palmstr{\o}m}
\affiliation{California NanoSystems Institute, University of California, Santa Barbara}
\affiliation{Electrical and Computer Engineering Department, University of California, Santa Barbara}
\affiliation{Materials Department, University of California, Santa Barbara}
\author{Vlad S. Pribiag}
	\email{vpribiag@umn.edu}
\affiliation{School of Physics and Astronomy, University of Minnesota}

\date{\today}

\begin{abstract} 
Josephson junctions with three or more superconducting leads have been predicted to exhibit topological effects in the presence of few conducting modes within the interstitial normal material. Such behavior, of relevance for topologically-protected quantum bits, would lead to specific transport features measured between terminals, with topological phase transitions occurring as a function of phase and voltage bias. Although conventional, two-terminal Josephson junctions have been studied extensively, multi-terminal devices have received relatively little attention to date. Motivated in part by the possibility to ultimately observe topological phenomena in multi-terminal Josephson devices, as well as their potential for coupling gatemon qubits, here we describe the superconducting features of a top-gated mesoscopic three-terminal Josephson device. The device is based on an InAs two-dimensional electron gas (2DEG) proximitized by epitaxial aluminum. We map out the transport properties of the device as a function of bias currents, top gate voltage and magnetic field. We find a very good agreement between the zero-field experimental phase diagram and a resistively and capacitively shunted junction (RCSJ) computational model.
\end{abstract}

\maketitle


\section{Introduction}

Superconductor-semiconductor-superconductor (S-Sm-S) junctions based on 1D and 2D semiconductors have recently attracted increasing attention, motivated by the possibility to realize novel phenomena enabled by the gate-control of induced superconductivity and by the interplay between superconductivity, spin-orbit coupling and topological boundary states \cite{Lutchyn2018, Doh2005, Gunel2012, Laroche2019, Kjaergaard2017, Suominen2017, Fornieri2019, Hart2014, Pribiag2015, Deacon2017, Ren2019, Williams2012, Ghatak2018}. In particular, two-dimensional electron gases (2DEGs) in semiconductor heterostructures have emerged as a promising platform for realizing gate-tunable S-Sm-S devices that can host topological states. The majority of work to date has focused on two-terminal Josephson junctions, where high interface transparency \cite{Kjaergaard2017, Shabani2016} and coherent ballistic transport \cite{Lee2019} have been demonstrated.  Experiments have also shown signatures of topological superconductivity in such junctions \cite{Hart2017,Ren2019,Fornieri2019}. Topological superconductivity is associated with Majorana zero modes (MZMs), which underpin proposals for fault tolerant topological quantum computation \cite{Kitaev2001,Kitaev2003,Nayak2008, Sarma2015}.

\begin{figure}[h!]
    \includegraphics[width=\linewidth]{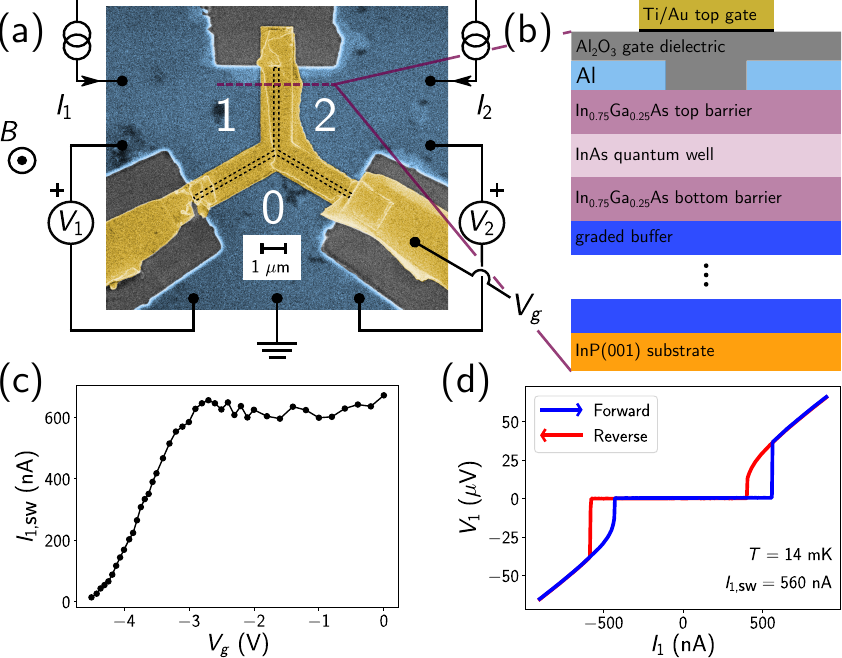}
    \caption{(a) False-color SEM image of gated three-terminal junction with measurement schematic. Blue areas are aluminum and grey areas are etched to the insulating buffer layers to create the device mesas. The Ti/Au top gate (yellow) overlays the 200 nm-wide Y-shaped junction formed by selectively etching the Al layer only (dotted lines). (b) Cross-sectional view along the purple horizontal line in (a) (not to scale). (c) Gate dependence of the switching current when biasing between terminals 1 and 2 showing pinch-off occurring at $V_g \sim -4.5$ V. (d) Two-terminal $I_1$ vs $V_1$ curve with $I_2=0$ showing hysteresis.}
    \label{fig1}
\end{figure}

A conventional two-terminal Josephson junction is described by a simple Josephson relation between the phase or voltage difference between the superconducting terminals \cite{Josephson1962}. By increasing the number of terminals, one can access a higher dimensional phase space spanned by the relative phases or voltages between the several terminals. This can lead to new effects, such as interactions between supercurrents \cite{Pfeffer2014,Cohen2018}, coexistence of dissipative currents and supercurrents \cite{Draelos2018}, multi-loop superconducting interferometry \cite{Vischi2017,Strambini2016}, multi-terminal Shapiro plateaus \cite{Deb2018}, or generalizations of multiple Andreev reflection \cite{Nowak2019,Pankratova2018}. Gated multi-terminal Josephson junctions have also been proposed\cite{Qi2018} as a means of coupling gatemon-type qubits\cite{Casparis2018}.

Several recent theoretical studies have proposed the existence of topological states in the Andreev spectrum of multi-terminal Josephson junctions, which under certain conditions could host zero-energy Weyl singularities \cite{Riwar2016,Meyer2017,Xie2017,Xie2018}. The topological nature of these states may protect them from conventional forms of quantum decoherence, a major hindrance to the advancement of robust and scalable quantum computation. These proposals consider multiple superconducting leads coupled to each other through a point-like central normal region which can be described by a single scattering matrix, $\hat{S}$, within which all pair-wise currents flow through a small number of modes. The topological phase transitions in the Andreev levels manifest quantized conductances and transconductances which change as a function of the terminal phases and/or voltages. 

In practice, the fabrication of S-2DEG-S junctions has focused primarily on devices with geometric extent that puts them far from the constraints present in the aforementioned theory work \cite{Hart2014, Pribiag2015, Kjaergaard2017}. That is, there are typically on the order of hundreds of current-carrying modes, and the regions in which scattering can occur are far from point-like. Nevertheless, as device designs and fabrication techniques improve, future devices may approach these proposed transport requirements. It is then important to characterize the background non-topological transport characteristics of a multi-terminal Josephson device using one of the most promising material platforms, an InAs quantum well proximitized by aluminum \cite{Shabani2016,Kjaergaard2017}.

\section{Gated Three-Terminal Josephson Junction}

Here we study a gated three-terminal Josephson device fabricated from an InAs quantum well heterostructure with a 10-nm epitaxial aluminum superconducting layer (Fig. \ref{fig1}(a)). The heterostructure was grown on an InP(001) substrate using molecular beam epitaxy. From the bottom up, it consists of an In$_{x}$Al$_{1-x}$As graded buffer (from $x=0.52$ to 0.81), 25-nm In$_{0.81}$Ga$_{0.19}$As/In$_{0.81}$Al$_{0.19}$As superlattice, 100-nm In$_{0.81}$Al$_{0.19}$As with Si $\delta$-doping ($2 \times 10^{12}$ cm$^{-2}$), 6-nm In$_{0.75}$Ga$_{0.25}$As bottom barrier, 7-nm InAs quantum well, and a 10-nm In$_{0.75}$Ga$_{0.25}$As top barrier (Fig. \ref{fig1}(b))\cite{Lee2019_2}. The sample has a measured carrier concentration of $n = 1.05\times 10^{12}$ cm$^{-2}$ and a mobility $\mu = 3.0\times 10^{4} $ cm$^2$/Vs. Standard electron-beam lithography (EBL) and wet etching were used to define an electrically isolated mesa, and to selectively etch the epitaxial Al into a 200-nm-wide Y-shaped junction in the central mesa area. Approximately 40 nm of Al$_2$O$_3$ dielectric was deposited uniformly over the device die by atomic layer deposition. A Ti/Au topgate was defined using EBL and deposited via electron-beam evaporation to cover the etched Y-shaped junction. We estimate the mean free path in the 2DEG to be $\ell \sim 500$ nm. This puts our device in an intermediate mesoscopic regime, where transport directly across the junction is expected to occur ballistically, however due to the randomly-distributed transmission coefficients of the many modes present (each junction arm is 4 $\mu$m long) the average transport properties of the device are expected to have features of the diffusive limit. Moreover, in the semi-classical picture, electrons entering the junction with non-zero momentum parallel to the contacts can travel distances longer than the mean free path, given the 4 $\mu$m length of the contacts, which again puts the devices in an intermediate regime.

We performed DC current-biased measurements in a dilution refrigerator with a base temperature of $T \sim 14$ mK. We label the current applied between terminals 1 and 0 as $I_1$, the current applied between 2 and 0 as $I_2$, and the current between 1 and 2 as $I_{12}$. By making the topgate voltage more negative, the electron density in the interstitial 2DEG is gradually depleted, which tunes the switching current from $\sim$560 nA to 0 (Fig. \ref{fig1}(c)). As shown in Fig. \ref{fig1}(d) the junctions exhibit hysteresis with respect to the current sweep direction. Hysteretic Josephson I-V curves can occur either due to the presence of a shunt capacitance (as described by the resistively and capacitively shunted junction model, RCSJ \cite{McCumber1968}) or due to Joule self-heating which sets on as the junction becomes dissipative \cite{Courtois2008}\cite{DeCecco2016}. Although 2DEG-based lateral Josephson junctions can in principle have large shunt capacitances due to conducting underlayers in the heterostructures or capacitive coupling of each superconducting terminal to the topgate, we estimate that the capacitance in our device is small (Stewart-McCumber parameter $\beta_c=(\hbar/2e)I_cR_n^2C \sim 0.08$), making it likely that the hysteresis originates predominantly from Joule heating rather than capacitive effects.

\section{Measurements}

\begin{figure}[t]
    \includegraphics[width=\linewidth]{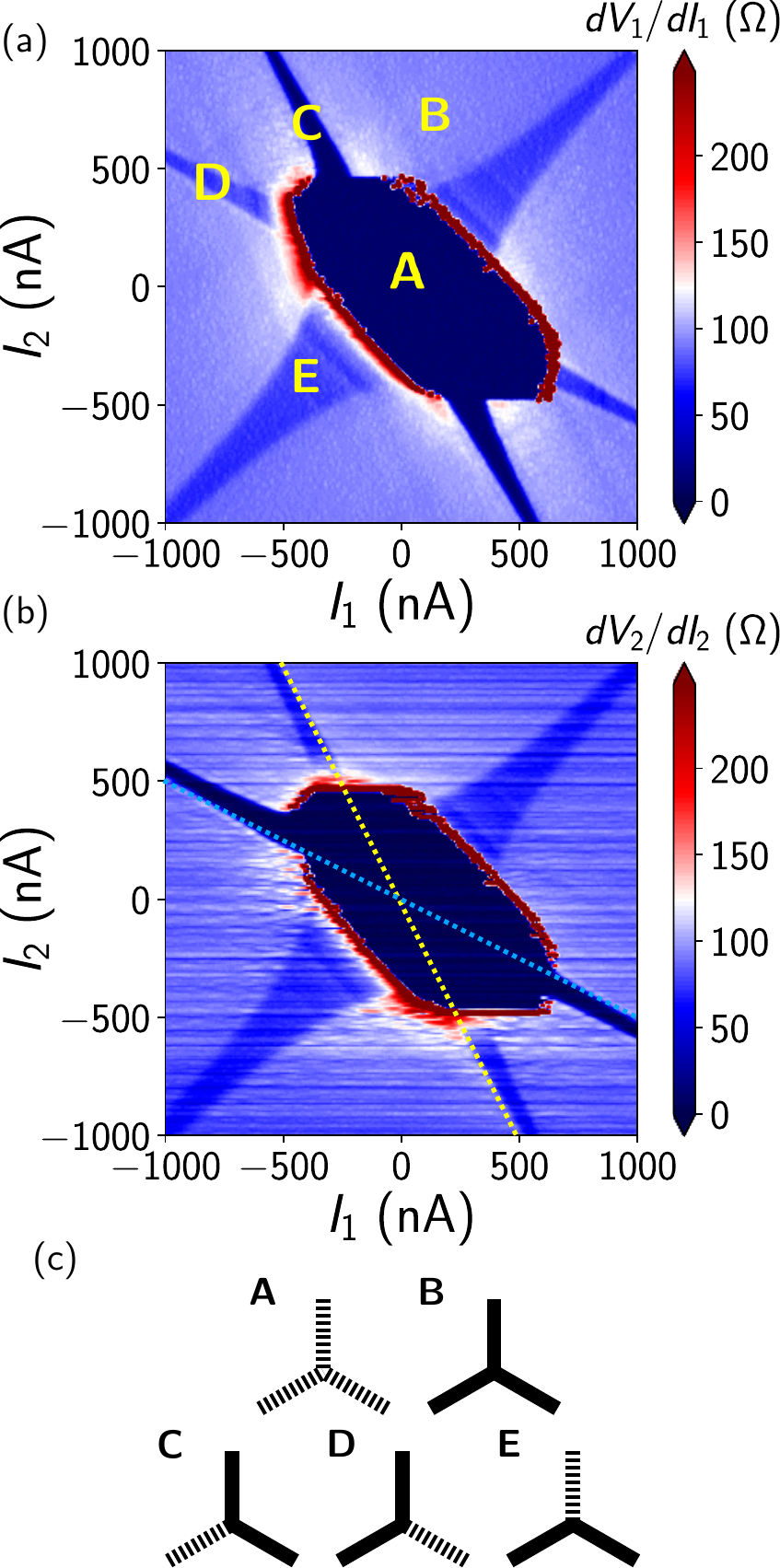}
    \caption{(a) $dV_1/dI_1$ as measured at $B = 0$, $V_g = -2$ V. The five distinct transport regions are indicated by the letters A-E. (b) $dV_2/dI_2$ from the same data as (a). The dotted yellow line indicates $I_2 = -2 I_1$, and the dotted blue line $I_1 = -2 I_2$. (c) Schematic depictions of the junction states corresponding to each of the lettered regimes from (a). Dashed (solid) lines indicate superconducting (resistive) legs of the tri-junction.}
    \label{fig2}
\end{figure}

To map out the behavior of the device, we performed three-terminal measurements by applying independent DC current-bias to terminals 1 ($I_1$) and 2 ($I_2$), with terminal 0 acting as the ground. We simultaneously measure the voltage of terminals 1 ($V_1$) and 2 ($V_2$) relative to terminal 0 (Fig. \ref{fig1}(a)). For the measurements included in this paper, we step $I_2$ from negative to positive and sweep $I_1$ from negative to positive at each value of $I_2$. A three-terminal data point then consists of a tuple ($I_1$, $I_2$, $V_1$, $V_2$). As the three-junctions are interconnected, the voltages are each functions of both input currents, $V_1 (I_1, I_2)$ and $V_2 (I_1,I_2)$. To visualize this data, we can discretely differentiate the voltages with respect to their corresponding current to get the differential resistances $dV_1/dI_1$ and $dV_2/dI_2$. To stabilize against any gate noise instabilities, we use a negative gate voltage of $V_g= -2$ V, which does not measurably affect the switching currents. 

The phase diagram of the device vs. $I_1$ and $I_2$ exhibits a central superconducting region where both $V_1$ and $V_2$ vanish, indicating that all three junctions carry supercurrent. At $V_g= -2$ V and with no magnetic field applied, this central region takes the shape of a rounded paralellogram (Figs. \ref{fig2}(a) and (b)).

Superconducting features extend beyond the central region A, and correspond to different combinations of dissipationless or resistive transport across the three legs of the device. We label the distinct regions of the phase diagram with the letters A-E in Fig. \ref{fig2}(a) and identify the junction configurations by corresponding schematic representations in Fig. \ref{fig2}(c). For example, regions C and D correspond to the relationships $I_2 = -2 I_1$ and $I_1 = -2 I_2$ respectively. Along region C we find $V_1 = 0$ and $dV_1/dI_1 = 0$, however $V_2$ and $dV_2/dI_2$ are nonzero. This indicates that there is a finite voltage difference between terminals 1 and 2 and thus no supercurrent flowing between them. Thus this arm corresponds to a region in current space where the junction formed by terminals 1 and 0 is carrying supercurrent, while the other two junctions are resistive and carry dissipative currents. To understand the factor of two relationship between $I_1$ and $I_2$ we consider the geometric current path that $I_2$ takes as it encounters two resistive junctions while the third junction (between terminals 1 and 0) carries supercurrent (Fig. \ref{fig2}(c), scenario C). A positive $I_2$ applied while the device is in this state will be divided into two equal components, assuming the resistances of the two legs of the tri-junction are identical. The component of $I_2$ travelling first to terminal 1 before reaching terminal 0 will add to $I_1$. Thus when $I_2 = -2 I_1$, there will be approximately zero net current between terminals 1 and 0. Small deviations about this line are also dissipationless as long as the critical current density in the region between terminals 1 and 0 is not exceeded, giving region C a finite width. A similar argument applies to region D. Note that region D becomes dissipationless on the map of $dV_2/dI_2$ vs. $I_1$ and $I_2$ (Fig. \ref{fig2}(b)), while region C acquires a finite differential resistance on this map. This reflection symmetry in the line $I_1=-I_2$ further confirms that the labelled regions correspond to the configurations depicted in Fig. \ref{fig2}(c).

A distinct region is centered on the line $I_1 = I_2$ (region E). This feature corresponds to the case when there is no voltage difference between the current-biased terminals 1 and 2, i.e. $V_1 = V_2$. In this regime the junction between terminals 1 and 2 carries supercurrent while the other two junctions are resistive. Applying a Y-$\Delta$ transformation, we expect the effective resistance between terminals 1 and 0 in region E of Fig. \ref{fig2}(a) to be $R_E/R_A = 0.75$ of that in region B. Here, $R_E$ and $R_A$ are the average measured differential resistances in regions E and A of Fig. \ref{fig1}(a). This is close to the measured values of 0.80-0.85. The small deviation can be accounted for by the fact that the resistances between terminals are not precisely equal. We find that the widths of the superconducting arms (regions C, D and E) shrink as the applied currents increase. We attribute this to decreased critical currents owing to Joule heating, which becomes more important at higher combined currents $I_1+I_2$ \cite{Draelos2018}.

\begin{figure}[]
    \includegraphics[width=\linewidth]{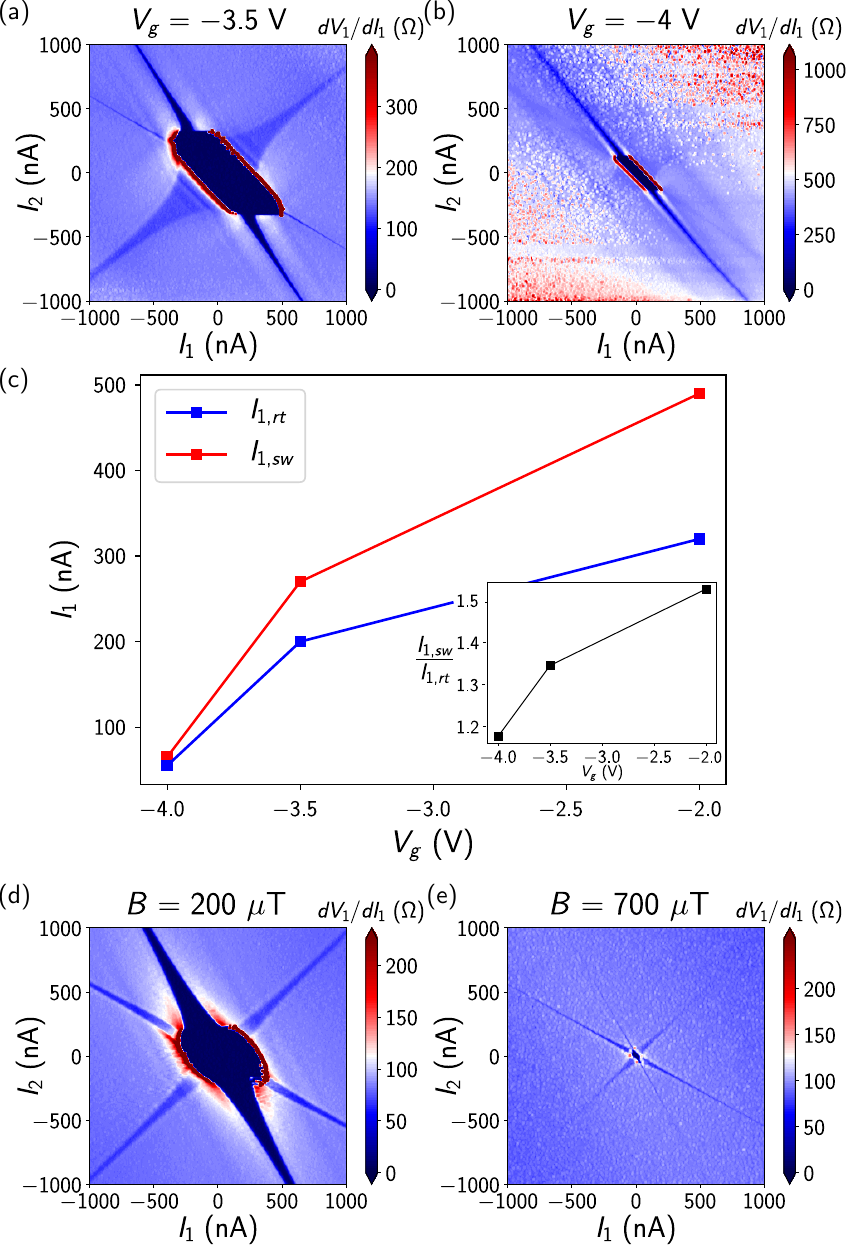}
    \caption{(a), (b) $dV_1/dI_1$ at two different gate voltages. (c) Dependence of retrapping current ($I_{1,\textrm{rt}}$) and switching current ($I_{1,\textrm{sw}}$) on gate voltage, with $I_2=0$. Inset shows the ratio $I_{1,\textrm{sw}}/I_{1,\textrm{rt}}$ vs. $V_g$. (d), (e) $dV_1/dI_1$ vs. $I_1$ and $I_2$ at two different values of the perpendicular magnetic field.} 
    \label{fig3}
\end{figure}

Applying more negative gate voltages reduces the extent of regions A and C-E and increases the resistances of regions C-E, as expected since the electron density in the InAs is decreased and the critical currents are consequently being reduced (Figs. \ref{fig3}(a) and (b)). We observe that the width of region D is reduced faster than that of region C, such that it becomes nearly unresolvable at $V_g = -4$ V. This indicates that the top gating, though nominally symmetric, has a slightly higher efficiency for the InAs region between terminals 2 and 0 than for that between terminals 1 and 0. In addition, the arms of regions C and D tilt away from the slopes described previously, which we also attribute to the differential effect of gating on the resistances of each junction. Interestingly, we also observe a gradual vanishing of the hysteresis for more negative $V_g$, as seen in Fig. \ref{fig3}(c). This is in agreement with a thermal origin of the observed hysteresis, since as the critical currents are reduced by gating, the Joule heating power is also reduced nearby these smaller switching currents. These interpretations are supported by detailed simulation results, presented in the next section. 

We have also explored the effect of a small perpendicular magnetic field, corresponding to less than one flux quantum through the total area of the junctions (Figs. \ref{fig3}(d) and (e)). As expected, the magnetic field does not affect the resistances of regions B-E. The effect of the field, like that of the topgate voltage, is to shrink all of the regions associated with superconductivity (A, C-E). However, in contrast with the asymmetric effect of $V_g$ noted above, applying a magnetic field causes all regions to shrink in roughly equal proportions, consistent with a spatially homogeneous weakening of superconductivity. This could arise from modulation of the critical currents due to superconducting quantum interference (analogous to the Fraunhofer-like modulation of $I_c$ in a lateral two-terminal Josephson junction \cite{Pribiag2015}) or due to dissipation by superconducting vortices in the thin-film aluminum leads \cite{Tinkham1963,Maki1965}. Another marked difference between the effects of magnetic field and $V_g$ pertains to the shape of the central superconducting region (A). While gating makes region A more parallelogram-like, a finite field makes it more elliptical in shape. We discuss the differences between gating and magnetic field in more detail below.

\section{Simulations}

\begin{figure*}[]
    \includegraphics[width=\linewidth]{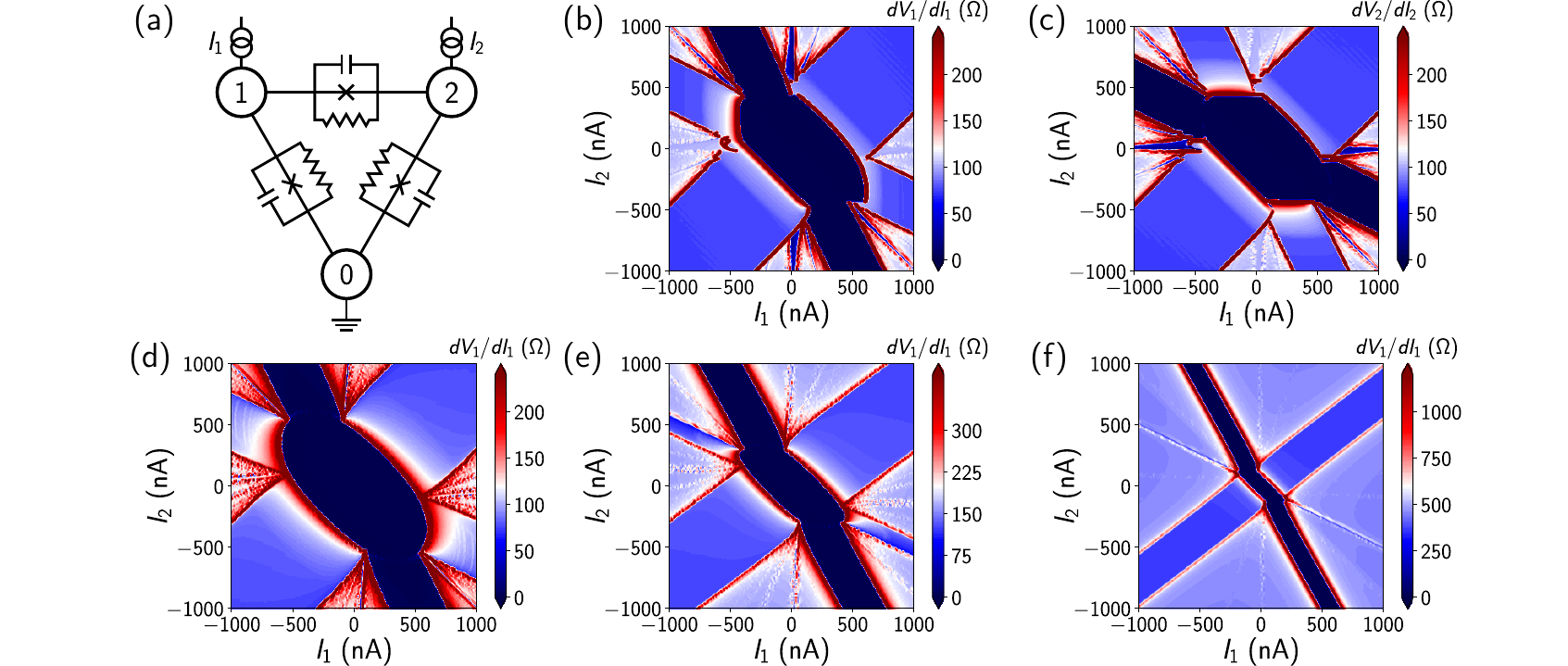}
    \caption{(a) Three-junction RCSJ network used in simulations. (b),(c) Simulated plots of $dV_1/dI_1$ and $dV_2/dI_2$ with large synthetic capacitance, which reproduce the main features in Fig. \ref{fig2}(a) and (b), respectively. There are additional very narrow superconducting arms along the lines $I_1 =0$, $I_2 =0$ and $I_1=-I_2$, which we do not see in the data. These features could be completely suppressed in the measurements by heating, in the same way that the experimentally-visible superconducting arms are narrowed. (d) Simulation with same parameters as (b) and (c), but with $C_i=0$, showing that the central region becomes elliptical when hysteresis is absent and the three critical currents are relatively close in magnitude. (e) $dV_1/dI_1$ with reduced $I_{c,i}$ to simulate gating effects (c.f. Fig. \ref{fig3}(a)). (f) Further reduction of critical currents compared to (e), c.f. Fig. \ref{fig3}(b). In (e) and (f) $I_{c,2}$ is reduced more than the other two critical currents to account for the observed gating asymmetry, and $C_i$ are set to zero to account for the small hysteresis observed in this regime (c.f. Figs. \ref{fig3}(a),(b) and (c)).}
    \label{fig4}
\end{figure*}

To gain more insight into the phase diagram of our device, we employ a numerical simulation. We use the SPICE-based superconducting circuit simulator PSCAN2 \cite{Polonsky1991}. Using this simulator framework, we model our device as a network of three junctions in the RCSJ description (Fig. \ref{fig4}(a)). This simple model contains nine parameters: the critical currents $I_{c,i}$, normal-state resistances $R_{n,i}$, and capacitances $C_i$ for each of the three Josephson junctions in the network. The precise values of these parameters as well as the PSCAN2 code used to generate the items in Fig. 4 can be found in the supplementary material of this work \cite{Supp}.

State-of-the-art RCSJ simulation software currently does not have the capability of including temperature effects due to Joule heating \cite{Fourie2018} and adding this capability is beyond the scope of this work. As a result, in order to capture the hysteretic effects observed in our device, we include instead a sufficiently large synthetic capacitance in each junction. We find that the key features of our experimental data from Fig. \ref{fig2} are all reproduced semi-quantitatively by our simulations (Fig. \ref{fig4}(b) and (c)). This includes the shape of the central superconducting region, the slope and position of the arms, and the resistance values in the dissipative regimes.  In contrast, removing the hysteresis by setting all $C_i=0$ (RSJ model) generally fails to reproduce the experimental central region, instead yielding an elliptical shape (Fig. \ref{fig4}(f)). On the basis of this very good overall agreement of our RCSJ simulations with the data, we conclude that modeling the hysteresis is necessary, but that its specific origin (capacitive or Joule heating) does not have a major impact on the phase diagram. As expected, the RCSJ simulations do not show the gradual tapering off of regions C-E, which occurs experimentally for larger applied currents, consistent with this effect being due to the dynamical reduction of the $I_{c,i}$ due to Joule heating.

The enhanced parallelogram shape of region A measured at more negative gate voltages (Figs. \ref{fig3}(a) and (b)) is also readily recovered in the simulations by assuming that gating decreases $I_{c,2}$ slightly more efficiently than $I_{c,1}$ and $I_{c,12}$ (Figs. \ref{fig4}(e) and (f)). This is consistent with our data showing the width of the superconducting arm along $I_1 = -2 I_2$ (region D) decreasing more rapidly relative to the other arms close to pinch-off due to the differential effect of gating on the three legs of the junction. Note also that in this regime the experimental hysteresis becomes very small (see Fig. \ref{fig3}(c)), again consistent with a heating origin of the hysteresis. In the simulations, we model this low $I_{c,i}$ regime with vanishingly small $C_i$.   

We find that the extent of the central superconducting region and the widths of the arms are dependent upon all of the $I_{c,i}$ values used in the simulations in a nonlinear fashion. For example, a line-cut through the central region along $I_2 = 0$ does not correspond to a switching current equal to $I_{c,1}$. This is due to supercurrent splitting along two paths to reach ground. For example, with $I_2=0$, we see a switch to the resistive state occurring at the value $I_1 = I_{c,1}+I_{c,12}$, since the current flows not only directly between terminals 1 and 0, but also to 0 through terminal 2. Therefore the condition to have purely dissipative transport in the device is that $I_{c,1}$ and $I_{c,12}$ must both be exceeded simultaneously. In the case of a parallelogram-shaped central region, this relationship can be extended to explain the location of the sides of the region. The upper and lower horizontal boundaries occur when $I_2 = \pm (I_{c,1} + I_{c,12})$. The angled sides then correspond to $I_1 + I_2 = \pm (I_{c,1} + I_{c,2})$, having a slope of minus one. This obviously does not hold for when the boundaries are more rounded, as are seen in experimental data under magnetic field, and also the simulation with $C_i=0$.

The discussion in the preceding paragraphs shows that at $B=0$ our three-terminal device can be very well understood using a model of three classically-coupled Josephson junctions with gate-tunable hysteresis. At finite $B$, the shape of the central region becomes more elliptical (Figs. \ref{fig3}(d) and (e)). This is qualitatively similar to the results of our simulations with $C_i=0$ and relatively symmetric $I_{c,i}$ values (Fig. \ref{fig4}(d)). Thus, the magnetic field may enhance the symmetry of the device, making the three $I_{c,i}$ values more similar, perhaps due to flux focusing \cite{Hart2014}. However, at finite magnetic field, the RCSJ model is fundamentally no longer applicable \cite{Altshuler2003}. Instead, a fully self-consistent theoretical treatment of the device, taking into account how the superconducting phases across the device vary with field, is required. The result for the case of two-terminal junctions is well known, especially for the SIS case, but the authors are not aware of any equivalent theory for multi-terminal junctions.

\section{Conclusions}

In conclusion, we report the behavior of a three-terminal Josephson junction based on an InAs quantum well with epitaxial aluminum superconducting leads as a function of electrostatic gating and applied perpendicular magnetic field. The ability to interpret and distinguish features that are due to mesoscopic superconducting transport in a multi-terminal Josephson junction is expected to prove useful in future studies which aim to approach the small-area, few-modes regime that is predicted to show topological effects. Such multi-terminal topological effects could provide a new path for the development of qubits with enhanced resilience to decoherence. Understanding the effects of gating and applied magnetic fields on a mesoscopic three-terminal Josephson junction could also help the development of coupled gatemon qubits. The combination of controlled gating, high mobility, and geometrically flexible fabrication present in this material platform makes it an excellent candidate for pursuing these goals.

\section{Acknowledgements}

We thank Alex Levchenko and Manuel Houzet for helpful discussions. This work was supported primarily by the National Science Foundation under Award No. DMR-1554609. The work at UCSB was supported by the Department of Energy under Award No. DE-SC0019274. The development of the epitaxial growth process was supported by Microsoft Research. Portions of this work were conducted  in the Minnesota Nano Center, which is supported by the National Science Foundation through the National Nano Coordinated Infrastructure Network (NNCI) under Award Number ECCS-1542202. 

\nocite{Nagel1973}


\bibliography{manuscript}

\providecommand{\noopsort}[1]{}\providecommand{\singleletter}[1]{#1}%
\begin{thebibliography}{46}%
\makeatletter
\providecommand \@ifxundefined [1]{%
 \@ifx{#1\undefined}
}%
\providecommand \@ifnum [1]{%
 \ifnum #1\expandafter \@firstoftwo
 \else \expandafter \@secondoftwo
 \fi
}%
\providecommand \@ifx [1]{%
 \ifx #1\expandafter \@firstoftwo
 \else \expandafter \@secondoftwo
 \fi
}%
\providecommand \natexlab [1]{#1}%
\providecommand \enquote  [1]{``#1''}%
\providecommand \bibnamefont  [1]{#1}%
\providecommand \bibfnamefont [1]{#1}%
\providecommand \citenamefont [1]{#1}%
\providecommand \href@noop [0]{\@secondoftwo}%
\providecommand \href [0]{\begingroup \@sanitize@url \@href}%
\providecommand \@href[1]{\@@startlink{#1}\@@href}%
\providecommand \@@href[1]{\endgroup#1\@@endlink}%
\providecommand \@sanitize@url [0]{\catcode `\\12\catcode `\$12\catcode
  `\&12\catcode `\#12\catcode `\^12\catcode `\_12\catcode `\%12\relax}%
\providecommand \@@startlink[1]{}%
\providecommand \@@endlink[0]{}%
\providecommand \url  [0]{\begingroup\@sanitize@url \@url }%
\providecommand \@url [1]{\endgroup\@href {#1}{\urlprefix }}%
\providecommand \urlprefix  [0]{URL }%
\providecommand \Eprint [0]{\href }%
\providecommand \doibase [0]{https://doi.org/}%
\providecommand \selectlanguage [0]{\@gobble}%
\providecommand \bibinfo  [0]{\@secondoftwo}%
\providecommand \bibfield  [0]{\@secondoftwo}%
\providecommand \translation [1]{[#1]}%
\providecommand \BibitemOpen [0]{}%
\providecommand \bibitemStop [0]{}%
\providecommand \bibitemNoStop [0]{.\EOS\space}%
\providecommand \EOS [0]{\spacefactor3000\relax}%
\providecommand \BibitemShut  [1]{\csname bibitem#1\endcsname}%
\let\auto@bib@innerbib\@empty
\bibitem [{\citenamefont {Lutchyn}\ \emph {et~al.}(2018)\citenamefont
  {Lutchyn}, \citenamefont {Bakkers}, \citenamefont {Kouwenhoven},
  \citenamefont {Krogstrup}, \citenamefont {Marcus},\ and\ \citenamefont
  {Oreg}}]{Lutchyn2018}%
  \BibitemOpen
  \bibfield  {author} {\bibinfo {author} {\bibfnamefont {R.~M.}\ \bibnamefont
  {Lutchyn}}, \bibinfo {author} {\bibfnamefont {E.~P.}\ \bibnamefont
  {Bakkers}}, \bibinfo {author} {\bibfnamefont {L.~P.}\ \bibnamefont
  {Kouwenhoven}}, \bibinfo {author} {\bibfnamefont {P.}~\bibnamefont
  {Krogstrup}}, \bibinfo {author} {\bibfnamefont {C.~M.}\ \bibnamefont
  {Marcus}}, and\ \bibinfo {author} {\bibfnamefont {Y.}~\bibnamefont {Oreg}},\
  }\bibfield  {title} {\bibinfo {title} {{Majorana zero modes in
  superconductor-semiconductor heterostructures}},\ }\href
  {https://doi.org/10.1038/s41578-018-0003-1} {\bibfield  {journal} {\bibinfo
  {journal} {Nat. Rev. Mater.}\ }\textbf {\bibinfo {volume} {3}},\ \bibinfo
  {pages} {52} (\bibinfo {year} {2018})}\BibitemShut {NoStop}%
\bibitem [{\citenamefont {Doh}\ \emph {et~al.}(2005)\citenamefont {Doh},
  \citenamefont {van Dam}, \citenamefont {Roest}, \citenamefont {Bakkers},
  \citenamefont {Kouwenhoven},\ and\ \citenamefont {{De
  Franceschi}}}]{Doh2005}%
  \BibitemOpen
  \bibfield  {author} {\bibinfo {author} {\bibfnamefont {Y.-J.}\ \bibnamefont
  {Doh}}, \bibinfo {author} {\bibfnamefont {J.~A.}\ \bibnamefont {van Dam}},
  \bibinfo {author} {\bibfnamefont {A.~L.}\ \bibnamefont {Roest}}, \bibinfo
  {author} {\bibfnamefont {E.~P. A.~M.}\ \bibnamefont {Bakkers}}, \bibinfo
  {author} {\bibfnamefont {L.~P.}\ \bibnamefont {Kouwenhoven}}, and\ \bibinfo
  {author} {\bibfnamefont {S.}~\bibnamefont {{De Franceschi}}},\ }\bibfield
  {title} {\bibinfo {title} {{Tunable supercurrent through semiconductor
  nanowires.}},\ }\href {https://doi.org/10.1126/science.1113523} {\bibfield
  {journal} {\bibinfo  {journal} {Science}\ }\textbf {\bibinfo {volume}
  {309}},\ \bibinfo {pages} {272} (\bibinfo {year} {2005})}\BibitemShut
  {NoStop}%
\bibitem [{\citenamefont {G{\"{u}}nel}\ \emph {et~al.}(2012)\citenamefont
  {G{\"{u}}nel}, \citenamefont {Batov}, \citenamefont {Hardtdegen},
  \citenamefont {Sladek}, \citenamefont {Winden}, \citenamefont {Weis},
  \citenamefont {Panaitov}, \citenamefont {Gr{\"{u}}tzmacher},\ and\
  \citenamefont {Sch{\"{a}}pers}}]{Gunel2012}%
  \BibitemOpen
  \bibfield  {author} {\bibinfo {author} {\bibfnamefont {H.~Y.}\ \bibnamefont
  {G{\"{u}}nel}}, \bibinfo {author} {\bibfnamefont {I.~E.}\ \bibnamefont
  {Batov}}, \bibinfo {author} {\bibfnamefont {H.}~\bibnamefont {Hardtdegen}},
  \bibinfo {author} {\bibfnamefont {K.}~\bibnamefont {Sladek}}, \bibinfo
  {author} {\bibfnamefont {A.}~\bibnamefont {Winden}}, \bibinfo {author}
  {\bibfnamefont {K.}~\bibnamefont {Weis}}, \bibinfo {author} {\bibfnamefont
  {G.}~\bibnamefont {Panaitov}}, \bibinfo {author} {\bibfnamefont
  {D.}~\bibnamefont {Gr{\"{u}}tzmacher}}, and\ \bibinfo {author} {\bibfnamefont
  {T.}~\bibnamefont {Sch{\"{a}}pers}},\ }\bibfield  {title} {\bibinfo {title}
  {{Supercurrent in Nb/InAs-nanowire/Nb Josephson junctions}},\ }\href
  {https://doi.org/10.1063/1.4745024} {\bibfield  {journal} {\bibinfo
  {journal} {J. Appl. Phys.}\ }\textbf {\bibinfo {volume} {112}},\ \bibinfo
  {pages} {034316} (\bibinfo {year} {2012})}\BibitemShut {NoStop}%
\bibitem [{\citenamefont {Laroche}\ \emph {et~al.}(2019)\citenamefont
  {Laroche}, \citenamefont {Bouman}, \citenamefont {van Woerkom}, \citenamefont
  {Proutski}, \citenamefont {Murthy}, \citenamefont {Pikulin}, \citenamefont
  {Nayak}, \citenamefont {van Gulik}, \citenamefont {Nyg{\aa}rd}, \citenamefont
  {Krogstrup}, \citenamefont {Kouwenhoven},\ and\ \citenamefont
  {Geresdi}}]{Laroche2019}%
  \BibitemOpen
  \bibfield  {author} {\bibinfo {author} {\bibfnamefont {D.}~\bibnamefont
  {Laroche}}, \bibinfo {author} {\bibfnamefont {D.}~\bibnamefont {Bouman}},
  \bibinfo {author} {\bibfnamefont {D.~J.}\ \bibnamefont {van Woerkom}},
  \bibinfo {author} {\bibfnamefont {A.}~\bibnamefont {Proutski}}, \bibinfo
  {author} {\bibfnamefont {C.}~\bibnamefont {Murthy}}, \bibinfo {author}
  {\bibfnamefont {D.~I.}\ \bibnamefont {Pikulin}}, \bibinfo {author}
  {\bibfnamefont {C.}~\bibnamefont {Nayak}}, \bibinfo {author} {\bibfnamefont
  {R.~J.}\ \bibnamefont {van Gulik}}, \bibinfo {author} {\bibfnamefont
  {J.}~\bibnamefont {Nyg{\aa}rd}}, \bibinfo {author} {\bibfnamefont
  {P.}~\bibnamefont {Krogstrup}}, \bibinfo {author} {\bibfnamefont {L.~P.}\
  \bibnamefont {Kouwenhoven}}, and\ \bibinfo {author} {\bibfnamefont
  {A.}~\bibnamefont {Geresdi}},\ }\bibfield  {title} {\bibinfo {title}
  {{Observation of the 4$\pi$-periodic Josephson effect in indium arsenide
  nanowires}},\ }\href {https://doi.org/10.1038/s41467-018-08161-2} {\bibfield
  {journal} {\bibinfo  {journal} {Nat. Commun.}\ }\textbf {\bibinfo {volume}
  {10}},\ \bibinfo {pages} {245} (\bibinfo {year} {2019})}\BibitemShut
  {NoStop}%
\bibitem [{\citenamefont {Kjaergaard}\ \emph {et~al.}(2017)\citenamefont
  {Kjaergaard}, \citenamefont {Suominen}, \citenamefont {Nowak}, \citenamefont
  {Akhmerov}, \citenamefont {Shabani}, \citenamefont {Palmstr{\o}m},
  \citenamefont {Nichele},\ and\ \citenamefont {Marcus}}]{Kjaergaard2017}%
  \BibitemOpen
  \bibfield  {author} {\bibinfo {author} {\bibfnamefont {M.}~\bibnamefont
  {Kjaergaard}}, \bibinfo {author} {\bibfnamefont {H.~J.}\ \bibnamefont
  {Suominen}}, \bibinfo {author} {\bibfnamefont {M.~P.}\ \bibnamefont {Nowak}},
  \bibinfo {author} {\bibfnamefont {A.~R.}\ \bibnamefont {Akhmerov}}, \bibinfo
  {author} {\bibfnamefont {J.}~\bibnamefont {Shabani}}, \bibinfo {author}
  {\bibfnamefont {C.~J.}\ \bibnamefont {Palmstr{\o}m}}, \bibinfo {author}
  {\bibfnamefont {F.}~\bibnamefont {Nichele}}, and\ \bibinfo {author}
  {\bibfnamefont {C.~M.}\ \bibnamefont {Marcus}},\ }\bibfield  {title}
  {\bibinfo {title} {{Transparent Semiconductor-Superconductor Interface and
  Induced Gap in an Epitaxial Heterostructure Josephson Junction}},\ }\href
  {https://doi.org/10.1103/PhysRevApplied.7.034029} {\bibfield  {journal}
  {\bibinfo  {journal} {Phys. Rev. Appl.}\ }\textbf {\bibinfo {volume} {7}},\
  \bibinfo {pages} {034029} (\bibinfo {year} {2017})}\BibitemShut {NoStop}%
\bibitem [{\citenamefont {Suominen}\ \emph {et~al.}(2017)\citenamefont
  {Suominen}, \citenamefont {Kjaergaard}, \citenamefont {Hamilton},
  \citenamefont {Shabani}, \citenamefont {Palmstr{\o}m}, \citenamefont
  {Marcus},\ and\ \citenamefont {Nichele}}]{Suominen2017}%
  \BibitemOpen
  \bibfield  {author} {\bibinfo {author} {\bibfnamefont {H.~J.}\ \bibnamefont
  {Suominen}}, \bibinfo {author} {\bibfnamefont {M.}~\bibnamefont
  {Kjaergaard}}, \bibinfo {author} {\bibfnamefont {A.~R.}\ \bibnamefont
  {Hamilton}}, \bibinfo {author} {\bibfnamefont {J.}~\bibnamefont {Shabani}},
  \bibinfo {author} {\bibfnamefont {C.~J.}\ \bibnamefont {Palmstr{\o}m}},
  \bibinfo {author} {\bibfnamefont {C.~M.}\ \bibnamefont {Marcus}}, and\
  \bibinfo {author} {\bibfnamefont {F.}~\bibnamefont {Nichele}},\ }\bibfield
  {title} {\bibinfo {title} {{Zero-Energy Modes from Coalescing Andreev States
  in a Two-Dimensional Semiconductor-Superconductor Hybrid Platform}},\ }\href
  {https://doi.org/10.1103/PhysRevLett.119.176805} {\bibfield  {journal}
  {\bibinfo  {journal} {Phys. Rev. Lett.}\ }\textbf {\bibinfo {volume} {119}},\
  \bibinfo {pages} {176805} (\bibinfo {year} {2017})}\BibitemShut {NoStop}%
\bibitem [{\citenamefont {Fornieri}\ \emph {et~al.}(2018)\citenamefont
  {Fornieri}, \citenamefont {Whiticar}, \citenamefont {Setiawan}, \citenamefont
  {Mar{\'{i}}n}, \citenamefont {Drachmann}, \citenamefont {Keselman},
  \citenamefont {Gronin}, \citenamefont {Thomas}, \citenamefont {Wang},
  \citenamefont {Kallaher}, \citenamefont {Gardner}, \citenamefont {Berg},
  \citenamefont {Manfra}, \citenamefont {Stern}, \citenamefont {Marcus},\ and\
  \citenamefont {Nichele}}]{Fornieri2019}%
  \BibitemOpen
  \bibfield  {author} {\bibinfo {author} {\bibfnamefont {A.}~\bibnamefont
  {Fornieri}}, \bibinfo {author} {\bibfnamefont {A.~M.}\ \bibnamefont
  {Whiticar}}, \bibinfo {author} {\bibfnamefont {F.}~\bibnamefont {Setiawan}},
  \bibinfo {author} {\bibfnamefont {E.~P.}\ \bibnamefont {Mar{\'{i}}n}},
  \bibinfo {author} {\bibfnamefont {A.~C.~C.}\ \bibnamefont {Drachmann}},
  \bibinfo {author} {\bibfnamefont {A.}~\bibnamefont {Keselman}}, \bibinfo
  {author} {\bibfnamefont {S.}~\bibnamefont {Gronin}}, \bibinfo {author}
  {\bibfnamefont {C.}~\bibnamefont {Thomas}}, \bibinfo {author} {\bibfnamefont
  {T.}~\bibnamefont {Wang}}, \bibinfo {author} {\bibfnamefont {R.}~\bibnamefont
  {Kallaher}}, \bibinfo {author} {\bibfnamefont {G.~C.}\ \bibnamefont
  {Gardner}}, \bibinfo {author} {\bibfnamefont {E.}~\bibnamefont {Berg}},
  \bibinfo {author} {\bibfnamefont {M.~J.}\ \bibnamefont {Manfra}}, \bibinfo
  {author} {\bibfnamefont {A.}~\bibnamefont {Stern}}, \bibinfo {author}
  {\bibfnamefont {C.~M.}\ \bibnamefont {Marcus}}, and\ \bibinfo {author}
  {\bibfnamefont {F.}~\bibnamefont {Nichele}},\ }\bibfield  {title} {\bibinfo
  {title} {{Evidence of topological superconductivity in planar Josephson
  junctions}},\ }\href {http://www.nature.com/articles/s41586-019-1068-8}
  {\bibfield  {journal} {\bibinfo  {journal} {Nature}\ }\textbf {\bibinfo
  {volume} {569}},\ \bibinfo {pages} {89} (\bibinfo {year} {2018})}\BibitemShut
  {NoStop}%
\bibitem [{\citenamefont {Hart}\ \emph {et~al.}(2014)\citenamefont {Hart},
  \citenamefont {Ren}, \citenamefont {Wagner}, \citenamefont {Leubner},
  \citenamefont {M{\"{u}}hlbauer}, \citenamefont {Br{\"{u}}ne}, \citenamefont
  {Buhmann}, \citenamefont {Molenkamp},\ and\ \citenamefont
  {Yacoby}}]{Hart2014}%
  \BibitemOpen
  \bibfield  {author} {\bibinfo {author} {\bibfnamefont {S.}~\bibnamefont
  {Hart}}, \bibinfo {author} {\bibfnamefont {H.}~\bibnamefont {Ren}}, \bibinfo
  {author} {\bibfnamefont {T.}~\bibnamefont {Wagner}}, \bibinfo {author}
  {\bibfnamefont {P.}~\bibnamefont {Leubner}}, \bibinfo {author} {\bibfnamefont
  {M.}~\bibnamefont {M{\"{u}}hlbauer}}, \bibinfo {author} {\bibfnamefont
  {C.}~\bibnamefont {Br{\"{u}}ne}}, \bibinfo {author} {\bibfnamefont
  {H.}~\bibnamefont {Buhmann}}, \bibinfo {author} {\bibfnamefont {L.~W.}\
  \bibnamefont {Molenkamp}}, and\ \bibinfo {author} {\bibfnamefont
  {A.}~\bibnamefont {Yacoby}},\ }\bibfield  {title} {\bibinfo {title} {{Induced
  superconductivity in the quantum spin Hall edge}},\ }\href
  {https://doi.org/10.1038/NPHYS3036} {\bibfield  {journal} {\bibinfo
  {journal} {Nat. Phys.}\ }\textbf {\bibinfo {volume} {10}},\ \bibinfo {pages}
  {638} (\bibinfo {year} {2014})}\BibitemShut {NoStop}%
\bibitem [{\citenamefont {Pribiag}\ \emph {et~al.}(2015)\citenamefont
  {Pribiag}, \citenamefont {Beukman}, \citenamefont {Qu}, \citenamefont
  {Cassidy}, \citenamefont {Charpentier}, \citenamefont {Wegscheider},\ and\
  \citenamefont {Kouwenhoven}}]{Pribiag2015}%
  \BibitemOpen
  \bibfield  {author} {\bibinfo {author} {\bibfnamefont {V.~S.}\ \bibnamefont
  {Pribiag}}, \bibinfo {author} {\bibfnamefont {A.~J.}\ \bibnamefont
  {Beukman}}, \bibinfo {author} {\bibfnamefont {F.}~\bibnamefont {Qu}},
  \bibinfo {author} {\bibfnamefont {M.~C.}\ \bibnamefont {Cassidy}}, \bibinfo
  {author} {\bibfnamefont {C.}~\bibnamefont {Charpentier}}, \bibinfo {author}
  {\bibfnamefont {W.}~\bibnamefont {Wegscheider}}, and\ \bibinfo {author}
  {\bibfnamefont {L.~P.}\ \bibnamefont {Kouwenhoven}},\ }\bibfield  {title}
  {\bibinfo {title} {{Edge-mode superconductivity in a two-dimensional
  topological insulator}},\ }\href {https://doi.org/10.1038/nnano.2015.86}
  {\bibfield  {journal} {\bibinfo  {journal} {Nat. Nanotechnol.}\ }\textbf
  {\bibinfo {volume} {10}},\ \bibinfo {pages} {593} (\bibinfo {year}
  {2015})}\BibitemShut {NoStop}%
\bibitem [{\citenamefont {Deacon}\ \emph {et~al.}(2017)\citenamefont {Deacon},
  \citenamefont {Wiedenmann}, \citenamefont {Bocquillon}, \citenamefont
  {Dom{\'{i}}nguez}, \citenamefont {Klapwijk}, \citenamefont {Leubner},
  \citenamefont {Br{\"{u}}ne}, \citenamefont {Hankiewicz}, \citenamefont
  {Tarucha}, \citenamefont {Ishibashi}, \citenamefont {Buhmann},\ and\
  \citenamefont {Molenkamp}}]{Deacon2017}%
  \BibitemOpen
  \bibfield  {author} {\bibinfo {author} {\bibfnamefont {R.~S.}\ \bibnamefont
  {Deacon}}, \bibinfo {author} {\bibfnamefont {J.}~\bibnamefont {Wiedenmann}},
  \bibinfo {author} {\bibfnamefont {E.}~\bibnamefont {Bocquillon}}, \bibinfo
  {author} {\bibfnamefont {F.}~\bibnamefont {Dom{\'{i}}nguez}}, \bibinfo
  {author} {\bibfnamefont {T.~M.}\ \bibnamefont {Klapwijk}}, \bibinfo {author}
  {\bibfnamefont {P.}~\bibnamefont {Leubner}}, \bibinfo {author} {\bibfnamefont
  {C.}~\bibnamefont {Br{\"{u}}ne}}, \bibinfo {author} {\bibfnamefont {E.~M.}\
  \bibnamefont {Hankiewicz}}, \bibinfo {author} {\bibfnamefont
  {S.}~\bibnamefont {Tarucha}}, \bibinfo {author} {\bibfnamefont
  {K.}~\bibnamefont {Ishibashi}}, \bibinfo {author} {\bibfnamefont
  {H.}~\bibnamefont {Buhmann}}, and\ \bibinfo {author} {\bibfnamefont {L.~W.}\
  \bibnamefont {Molenkamp}},\ }\bibfield  {title} {\bibinfo {title} {{Josephson
  radiation from gapless andreev bound states in HgTe-based topological
  junctions}},\ }\href {https://doi.org/10.1103/PhysRevX.7.021011} {\bibfield
  {journal} {\bibinfo  {journal} {Phys. Rev. X}\ }\textbf {\bibinfo {volume}
  {7}},\ \bibinfo {pages} {021011} (\bibinfo {year} {2017})}\BibitemShut
  {NoStop}%
\bibitem [{\citenamefont {Ren}\ \emph {et~al.}(2019)\citenamefont {Ren},
  \citenamefont {Pientka}, \citenamefont {Hart}, \citenamefont {Pierce},
  \citenamefont {Kosowsky}, \citenamefont {Lunczer}, \citenamefont {Schlereth},
  \citenamefont {Scharf}, \citenamefont {Hankiewicz}, \citenamefont
  {Molenkamp}, \citenamefont {Halperin},\ and\ \citenamefont
  {Yacoby}}]{Ren2019}%
  \BibitemOpen
  \bibfield  {author} {\bibinfo {author} {\bibfnamefont {H.}~\bibnamefont
  {Ren}}, \bibinfo {author} {\bibfnamefont {F.}~\bibnamefont {Pientka}},
  \bibinfo {author} {\bibfnamefont {S.}~\bibnamefont {Hart}}, \bibinfo {author}
  {\bibfnamefont {A.~T.}\ \bibnamefont {Pierce}}, \bibinfo {author}
  {\bibfnamefont {M.}~\bibnamefont {Kosowsky}}, \bibinfo {author}
  {\bibfnamefont {L.}~\bibnamefont {Lunczer}}, \bibinfo {author} {\bibfnamefont
  {R.}~\bibnamefont {Schlereth}}, \bibinfo {author} {\bibfnamefont
  {B.}~\bibnamefont {Scharf}}, \bibinfo {author} {\bibfnamefont {E.~M.}\
  \bibnamefont {Hankiewicz}}, \bibinfo {author} {\bibfnamefont {L.~W.}\
  \bibnamefont {Molenkamp}}, \bibinfo {author} {\bibfnamefont {B.~I.}\
  \bibnamefont {Halperin}}, and\ \bibinfo {author} {\bibfnamefont
  {A.}~\bibnamefont {Yacoby}},\ }\bibfield  {title} {\bibinfo {title}
  {{Topological superconductivity in a phase-controlled Josephson junction}},\
  }\href {https://www.nature.com/articles/s41586-019-1148-9} {\bibfield
  {journal} {\bibinfo  {journal} {Nature}\ }\textbf {\bibinfo {volume} {569}},\
  \bibinfo {pages} {93} (\bibinfo {year} {2019})}\BibitemShut {NoStop}%
\bibitem [{\citenamefont {Williams}\ \emph {et~al.}(2012)\citenamefont
  {Williams}, \citenamefont {Bestwick}, \citenamefont {Gallagher},
  \citenamefont {Hong}, \citenamefont {Cui}, \citenamefont {Bleich},
  \citenamefont {Analytis}, \citenamefont {Fisher},\ and\ \citenamefont
  {Goldhaber-Gordon}}]{Williams2012}%
  \BibitemOpen
  \bibfield  {author} {\bibinfo {author} {\bibfnamefont {J.~R.}\ \bibnamefont
  {Williams}}, \bibinfo {author} {\bibfnamefont {A.~J.}\ \bibnamefont
  {Bestwick}}, \bibinfo {author} {\bibfnamefont {P.}~\bibnamefont {Gallagher}},
  \bibinfo {author} {\bibfnamefont {S.~S.}\ \bibnamefont {Hong}}, \bibinfo
  {author} {\bibfnamefont {Y.}~\bibnamefont {Cui}}, \bibinfo {author}
  {\bibfnamefont {A.~S.}\ \bibnamefont {Bleich}}, \bibinfo {author}
  {\bibfnamefont {J.~G.}\ \bibnamefont {Analytis}}, \bibinfo {author}
  {\bibfnamefont {I.~R.}\ \bibnamefont {Fisher}}, and\ \bibinfo {author}
  {\bibfnamefont {D.}~\bibnamefont {Goldhaber-Gordon}},\ }\bibfield  {title}
  {\bibinfo {title} {{Unconventional Josephson Effect in Hybrid
  Superconductor-Topological Insulator Devices}},\ }\href
  {https://doi.org/10.1103/PhysRevLett.109.056803} {\bibfield  {journal}
  {\bibinfo  {journal} {Phys. Rev. Lett.}\ }\textbf {\bibinfo {volume} {109}},\
  \bibinfo {pages} {056803} (\bibinfo {year} {2012})}\BibitemShut {NoStop}%
\bibitem [{\citenamefont {Ghatak}\ \emph {et~al.}(2018)\citenamefont {Ghatak},
  \citenamefont {Breunig}, \citenamefont {Yang}, \citenamefont {Wang},
  \citenamefont {Taskin},\ and\ \citenamefont {Ando}}]{Ghatak2018}%
  \BibitemOpen
  \bibfield  {author} {\bibinfo {author} {\bibfnamefont {S.}~\bibnamefont
  {Ghatak}}, \bibinfo {author} {\bibfnamefont {O.}~\bibnamefont {Breunig}},
  \bibinfo {author} {\bibfnamefont {F.}~\bibnamefont {Yang}}, \bibinfo {author}
  {\bibfnamefont {Z.}~\bibnamefont {Wang}}, \bibinfo {author} {\bibfnamefont
  {A.~A.}\ \bibnamefont {Taskin}}, and\ \bibinfo {author} {\bibfnamefont
  {Y.}~\bibnamefont {Ando}},\ }\bibfield  {title} {\bibinfo {title} {{Anomalous
  Fraunhofer Patterns in Gated Josephson Junctions Based on the Bulk-Insulating
  Topological Insulator BiSbTeSe$_2$}},\ }\href
  {https://doi.org/10.1021/acs.nanolett.8b02029} {\bibfield  {journal}
  {\bibinfo  {journal} {Nano Lett.}\ }\textbf {\bibinfo {volume} {18}},\
  \bibinfo {pages} {5124} (\bibinfo {year} {2018})}\BibitemShut {NoStop}%
\bibitem [{\citenamefont {Shabani}\ \emph {et~al.}(2016)\citenamefont
  {Shabani}, \citenamefont {Kjaergaard}, \citenamefont {Suominen},
  \citenamefont {Kim}, \citenamefont {Nichele}, \citenamefont {Pakrouski},
  \citenamefont {Stankevic}, \citenamefont {Lutchyn}, \citenamefont
  {Krogstrup}, \citenamefont {Feidenhans'L}, \citenamefont {Kraemer},
  \citenamefont {Nayak}, \citenamefont {Troyer}, \citenamefont {Marcus},\ and\
  \citenamefont {Palmstr{\o}m}}]{Shabani2016}%
  \BibitemOpen
  \bibfield  {author} {\bibinfo {author} {\bibfnamefont {J.}~\bibnamefont
  {Shabani}}, \bibinfo {author} {\bibfnamefont {M.}~\bibnamefont {Kjaergaard}},
  \bibinfo {author} {\bibfnamefont {H.~J.}\ \bibnamefont {Suominen}}, \bibinfo
  {author} {\bibfnamefont {Y.}~\bibnamefont {Kim}}, \bibinfo {author}
  {\bibfnamefont {F.}~\bibnamefont {Nichele}}, \bibinfo {author} {\bibfnamefont
  {K.}~\bibnamefont {Pakrouski}}, \bibinfo {author} {\bibfnamefont
  {T.}~\bibnamefont {Stankevic}}, \bibinfo {author} {\bibfnamefont {R.~M.}\
  \bibnamefont {Lutchyn}}, \bibinfo {author} {\bibfnamefont {P.}~\bibnamefont
  {Krogstrup}}, \bibinfo {author} {\bibfnamefont {R.}~\bibnamefont
  {Feidenhans'L}}, \bibinfo {author} {\bibfnamefont {S.}~\bibnamefont
  {Kraemer}}, \bibinfo {author} {\bibfnamefont {C.}~\bibnamefont {Nayak}},
  \bibinfo {author} {\bibfnamefont {M.}~\bibnamefont {Troyer}}, \bibinfo
  {author} {\bibfnamefont {C.~M.}\ \bibnamefont {Marcus}}, and\ \bibinfo
  {author} {\bibfnamefont {C.~J.}\ \bibnamefont {Palmstr{\o}m}},\ }\bibfield
  {title} {\bibinfo {title} {{Two-dimensional epitaxial
  superconductor-semiconductor heterostructures: A platform for topological
  superconducting networks}},\ }\href
  {https://doi.org/10.1103/PhysRevB.93.155402} {\bibfield  {journal} {\bibinfo
  {journal} {Phys. Rev. B}\ }\textbf {\bibinfo {volume} {93}},\ \bibinfo
  {pages} {155402} (\bibinfo {year} {2016})}\BibitemShut {NoStop}%
\bibitem [{\citenamefont {Lee}\ \emph {et~al.}(2019{\natexlab{a}})\citenamefont
  {Lee}, \citenamefont {Shojaei}, \citenamefont {Pendharkar}, \citenamefont
  {McFadden}, \citenamefont {Kim}, \citenamefont {Suominen}, \citenamefont
  {Kjaergaard}, \citenamefont {Nichele}, \citenamefont {Zhang}, \citenamefont
  {Marcus},\ and\ \citenamefont {Palmstr{\o}m}}]{Lee2019}%
  \BibitemOpen
  \bibfield  {author} {\bibinfo {author} {\bibfnamefont {J.~S.}\ \bibnamefont
  {Lee}}, \bibinfo {author} {\bibfnamefont {B.}~\bibnamefont {Shojaei}},
  \bibinfo {author} {\bibfnamefont {M.}~\bibnamefont {Pendharkar}}, \bibinfo
  {author} {\bibfnamefont {A.~P.}\ \bibnamefont {McFadden}}, \bibinfo {author}
  {\bibfnamefont {Y.}~\bibnamefont {Kim}}, \bibinfo {author} {\bibfnamefont
  {H.~J.}\ \bibnamefont {Suominen}}, \bibinfo {author} {\bibfnamefont
  {M.}~\bibnamefont {Kjaergaard}}, \bibinfo {author} {\bibfnamefont
  {F.}~\bibnamefont {Nichele}}, \bibinfo {author} {\bibfnamefont
  {H.}~\bibnamefont {Zhang}}, \bibinfo {author} {\bibfnamefont {C.~M.}\
  \bibnamefont {Marcus}}, and\ \bibinfo {author} {\bibfnamefont {C.~J.}\
  \bibnamefont {Palmstr{\o}m}},\ }\bibfield  {title} {\bibinfo {title}
  {{Transport Studies of Epi-Al/InAs Two-Dimensional Electron Gas Systems for
  Required Building-Blocks in Topological Superconductor Networks}},\ }\href
  {https://doi.org/10.1021/acs.nanolett.9b00494} {\bibfield  {journal}
  {\bibinfo  {journal} {Nano Lett.}\ }\textbf {\bibinfo {volume} {19}},\
  \bibinfo {pages} {3083} (\bibinfo {year} {2019}{\natexlab{a}})}\BibitemShut
  {NoStop}%
\bibitem [{\citenamefont {Hart}\ \emph {et~al.}(2017)\citenamefont {Hart},
  \citenamefont {Ren}, \citenamefont {Kosowsky}, \citenamefont {Ben-Shach},
  \citenamefont {Leubner}, \citenamefont {Br{\"{u}}ne}, \citenamefont
  {Buhmann}, \citenamefont {Molenkamp}, \citenamefont {Halperin},\ and\
  \citenamefont {Yacoby}}]{Hart2017}%
  \BibitemOpen
  \bibfield  {author} {\bibinfo {author} {\bibfnamefont {S.}~\bibnamefont
  {Hart}}, \bibinfo {author} {\bibfnamefont {H.}~\bibnamefont {Ren}}, \bibinfo
  {author} {\bibfnamefont {M.}~\bibnamefont {Kosowsky}}, \bibinfo {author}
  {\bibfnamefont {G.}~\bibnamefont {Ben-Shach}}, \bibinfo {author}
  {\bibfnamefont {P.}~\bibnamefont {Leubner}}, \bibinfo {author} {\bibfnamefont
  {C.}~\bibnamefont {Br{\"{u}}ne}}, \bibinfo {author} {\bibfnamefont
  {H.}~\bibnamefont {Buhmann}}, \bibinfo {author} {\bibfnamefont {L.~W.}\
  \bibnamefont {Molenkamp}}, \bibinfo {author} {\bibfnamefont {B.~I.}\
  \bibnamefont {Halperin}}, and\ \bibinfo {author} {\bibfnamefont
  {A.}~\bibnamefont {Yacoby}},\ }\bibfield  {title} {\bibinfo {title}
  {{Controlled finite momentum pairing and spatially varying order parameter in
  proximitized HgTe quantum wells}},\ }\href
  {https://doi.org/10.1038/nphys3877} {\bibfield  {journal} {\bibinfo
  {journal} {Nat. Phys.}\ }\textbf {\bibinfo {volume} {13}},\ \bibinfo {pages}
  {87} (\bibinfo {year} {2017})}\BibitemShut {NoStop}%
\bibitem [{\citenamefont {Kitaev}(2001)}]{Kitaev2001}%
  \BibitemOpen
  \bibfield  {author} {\bibinfo {author} {\bibfnamefont {A.~Y.}\ \bibnamefont
  {Kitaev}},\ }\bibfield  {title} {\bibinfo {title} {{Unpaired Majorana
  fermions in quantum wires}},\ }\href
  {https://doi.org/10.1070/1063-7869/44/10S/S29} {\bibfield  {journal}
  {\bibinfo  {journal} {Phys.-Uspekhi}\ }\textbf {\bibinfo {volume} {44}},\
  \bibinfo {pages} {131} (\bibinfo {year} {2001})}\BibitemShut {NoStop}%
\bibitem [{\citenamefont {Kitaev}(2003)}]{Kitaev2003}%
  \BibitemOpen
  \bibfield  {author} {\bibinfo {author} {\bibfnamefont {A.~Y.}\ \bibnamefont
  {Kitaev}},\ }\bibfield  {title} {\bibinfo {title} {{Fault-tolerant quantum
  computation by anyons}},\ }\href
  {https://doi.org/10.1016/S0003-4916(02)00018-0} {\bibfield  {journal}
  {\bibinfo  {journal} {Ann. Phys. (N.Y.)}\ }\textbf {\bibinfo {volume}
  {303}},\ \bibinfo {pages} {2} (\bibinfo {year} {2003})}\BibitemShut {NoStop}%
\bibitem [{\citenamefont {Nayak}\ \emph {et~al.}(2008)\citenamefont {Nayak},
  \citenamefont {Simon}, \citenamefont {Stern}, \citenamefont {Freedman},\ and\
  \citenamefont {{Das Sarma}}}]{Nayak2008}%
  \BibitemOpen
  \bibfield  {author} {\bibinfo {author} {\bibfnamefont {C.}~\bibnamefont
  {Nayak}}, \bibinfo {author} {\bibfnamefont {S.~H.}\ \bibnamefont {Simon}},
  \bibinfo {author} {\bibfnamefont {A.}~\bibnamefont {Stern}}, \bibinfo
  {author} {\bibfnamefont {M.}~\bibnamefont {Freedman}}, and\ \bibinfo {author}
  {\bibfnamefont {S.}~\bibnamefont {{Das Sarma}}},\ }\bibfield  {title}
  {\bibinfo {title} {{Non-Abelian anyons and topological quantum
  computation}},\ }\href {https://doi.org/10.1103/RevModPhys.80.1083}
  {\bibfield  {journal} {\bibinfo  {journal} {Rev. Mod. Phys.}\ }\textbf
  {\bibinfo {volume} {80}},\ \bibinfo {pages} {1083} (\bibinfo {year}
  {2008})}\BibitemShut {NoStop}%
\bibitem [{\citenamefont {Sarma}\ \emph {et~al.}(2015)\citenamefont {Sarma},
  \citenamefont {Freedman},\ and\ \citenamefont {Nayak}}]{Sarma2015}%
  \BibitemOpen
  \bibfield  {author} {\bibinfo {author} {\bibfnamefont {S.~D.}\ \bibnamefont
  {Sarma}}, \bibinfo {author} {\bibfnamefont {M.}~\bibnamefont {Freedman}},
  and\ \bibinfo {author} {\bibfnamefont {C.}~\bibnamefont {Nayak}},\ }\bibfield
   {title} {\bibinfo {title} {{Majorana zero modes and topological quantum
  computation}},\ }\href {https://doi.org/10.1038/npjqi.2015.1} {\bibfield
  {journal} {\bibinfo  {journal} {npj Quantum Inf.}\ }\textbf {\bibinfo
  {volume} {1}},\ \bibinfo {pages} {15001} (\bibinfo {year}
  {2015})}\BibitemShut {NoStop}%
\bibitem [{\citenamefont {Josephson}(1962)}]{Josephson1962}%
  \BibitemOpen
  \bibfield  {author} {\bibinfo {author} {\bibfnamefont {B.}~\bibnamefont
  {Josephson}},\ }\bibfield  {title} {\bibinfo {title} {{Possible new effects
  in superconductive tunnelling}},\ }\href
  {https://doi.org/10.1016/0031-9163(62)91369-0} {\bibfield  {journal}
  {\bibinfo  {journal} {Phys. Lett.}\ }\textbf {\bibinfo {volume} {1}},\
  \bibinfo {pages} {251} (\bibinfo {year} {1962})}\BibitemShut {NoStop}%
\bibitem [{\citenamefont {Pfeffer}\ \emph {et~al.}(2014)\citenamefont
  {Pfeffer}, \citenamefont {Duvauchelle}, \citenamefont {Courtois},
  \citenamefont {M{\'{e}}lin}, \citenamefont {Feinberg},\ and\ \citenamefont
  {Lefloch}}]{Pfeffer2014}%
  \BibitemOpen
  \bibfield  {author} {\bibinfo {author} {\bibfnamefont {A.~H.}\ \bibnamefont
  {Pfeffer}}, \bibinfo {author} {\bibfnamefont {J.~E.}\ \bibnamefont
  {Duvauchelle}}, \bibinfo {author} {\bibfnamefont {H.}~\bibnamefont
  {Courtois}}, \bibinfo {author} {\bibfnamefont {R.}~\bibnamefont
  {M{\'{e}}lin}}, \bibinfo {author} {\bibfnamefont {D.}~\bibnamefont
  {Feinberg}}, and\ \bibinfo {author} {\bibfnamefont {F.}~\bibnamefont
  {Lefloch}},\ }\bibfield  {title} {\bibinfo {title} {{Subgap structure in the
  conductance of a three-terminal Josephson junction}},\ }\href
  {https://doi.org/10.1103/PhysRevB.90.075401} {\bibfield  {journal} {\bibinfo
  {journal} {Phys. Rev. B}\ }\textbf {\bibinfo {volume} {90}},\ \bibinfo
  {pages} {075401} (\bibinfo {year} {2014})}\BibitemShut {NoStop}%
\bibitem [{\citenamefont {Cohen}\ \emph {et~al.}(2018)\citenamefont {Cohen},
  \citenamefont {Ronen}, \citenamefont {Kang}, \citenamefont {Heiblum},
  \citenamefont {Feinberg}, \citenamefont {M{\'{e}}lin},\ and\ \citenamefont
  {Shtrikman}}]{Cohen2018}%
  \BibitemOpen
  \bibfield  {author} {\bibinfo {author} {\bibfnamefont {Y.}~\bibnamefont
  {Cohen}}, \bibinfo {author} {\bibfnamefont {Y.}~\bibnamefont {Ronen}},
  \bibinfo {author} {\bibfnamefont {J.-H.}\ \bibnamefont {Kang}}, \bibinfo
  {author} {\bibfnamefont {M.}~\bibnamefont {Heiblum}}, \bibinfo {author}
  {\bibfnamefont {D.}~\bibnamefont {Feinberg}}, \bibinfo {author}
  {\bibfnamefont {R.}~\bibnamefont {M{\'{e}}lin}}, and\ \bibinfo {author}
  {\bibfnamefont {H.}~\bibnamefont {Shtrikman}},\ }\bibfield  {title} {\bibinfo
  {title} {{Nonlocal supercurrent of quartets in a three-terminal Josephson
  junction}},\ }\href {https://doi.org/10.1073/pnas.1800044115} {\bibfield
  {journal} {\bibinfo  {journal} {Proc. Natl. Acad. Sci. U.S.A.}\ }\textbf
  {\bibinfo {volume} {115}},\ \bibinfo {pages} {6991} (\bibinfo {year}
  {2018})}\BibitemShut {NoStop}%
\bibitem [{\citenamefont {Draelos}\ \emph {et~al.}(2019)\citenamefont
  {Draelos}, \citenamefont {Wei}, \citenamefont {Seredinski}, \citenamefont
  {Li}, \citenamefont {Mehta}, \citenamefont {Watanabe}, \citenamefont
  {Taniguchi}, \citenamefont {Borzenets}, \citenamefont {Amet},\ and\
  \citenamefont {Finkelstein}}]{Draelos2018}%
  \BibitemOpen
  \bibfield  {author} {\bibinfo {author} {\bibfnamefont {A.~W.}\ \bibnamefont
  {Draelos}}, \bibinfo {author} {\bibfnamefont {M.-T.}\ \bibnamefont {Wei}},
  \bibinfo {author} {\bibfnamefont {A.}~\bibnamefont {Seredinski}}, \bibinfo
  {author} {\bibfnamefont {H.}~\bibnamefont {Li}}, \bibinfo {author}
  {\bibfnamefont {Y.}~\bibnamefont {Mehta}}, \bibinfo {author} {\bibfnamefont
  {K.}~\bibnamefont {Watanabe}}, \bibinfo {author} {\bibfnamefont
  {T.}~\bibnamefont {Taniguchi}}, \bibinfo {author} {\bibfnamefont {I.~V.}\
  \bibnamefont {Borzenets}}, \bibinfo {author} {\bibfnamefont {F.}~\bibnamefont
  {Amet}}, and\ \bibinfo {author} {\bibfnamefont {G.}~\bibnamefont
  {Finkelstein}},\ }\bibfield  {title} {\bibinfo {title} {{Supercurrent Flow in
  Multiterminal Graphene Josephson Junctions}},\ }\href
  {https://doi.org/10.1021/acs.nanolett.8b04330} {\bibfield  {journal}
  {\bibinfo  {journal} {Nano Lett.}\ }\textbf {\bibinfo {volume} {19}},\
  \bibinfo {pages} {1039} (\bibinfo {year} {2019})}\BibitemShut {NoStop}%
\bibitem [{\citenamefont {Vischi}\ \emph {et~al.}(2017)\citenamefont {Vischi},
  \citenamefont {Carrega}, \citenamefont {Strambini}, \citenamefont
  {D'Ambrosio}, \citenamefont {Bergeret}, \citenamefont {Nazarov},\ and\
  \citenamefont {Giazotto}}]{Vischi2017}%
  \BibitemOpen
  \bibfield  {author} {\bibinfo {author} {\bibfnamefont {F.}~\bibnamefont
  {Vischi}}, \bibinfo {author} {\bibfnamefont {M.}~\bibnamefont {Carrega}},
  \bibinfo {author} {\bibfnamefont {E.}~\bibnamefont {Strambini}}, \bibinfo
  {author} {\bibfnamefont {S.}~\bibnamefont {D'Ambrosio}}, \bibinfo {author}
  {\bibfnamefont {F.~S.}\ \bibnamefont {Bergeret}}, \bibinfo {author}
  {\bibfnamefont {Y.~V.}\ \bibnamefont {Nazarov}}, and\ \bibinfo {author}
  {\bibfnamefont {F.}~\bibnamefont {Giazotto}},\ }\bibfield  {title} {\bibinfo
  {title} {{Coherent transport properties of a three-terminal hybrid
  superconducting interferometer}},\ }\href
  {https://doi.org/10.1103/PhysRevB.95.054504} {\bibfield  {journal} {\bibinfo
  {journal} {Phys. Rev. B}\ }\textbf {\bibinfo {volume} {95}},\ \bibinfo
  {pages} {054504} (\bibinfo {year} {2017})}\BibitemShut {NoStop}%
\bibitem [{\citenamefont {Strambini}\ \emph {et~al.}(2016)\citenamefont
  {Strambini}, \citenamefont {D'Ambrosio}, \citenamefont {Vischi},
  \citenamefont {Bergeret}, \citenamefont {Nazarov},\ and\ \citenamefont
  {Giazotto}}]{Strambini2016}%
  \BibitemOpen
  \bibfield  {author} {\bibinfo {author} {\bibfnamefont {E.}~\bibnamefont
  {Strambini}}, \bibinfo {author} {\bibfnamefont {S.}~\bibnamefont
  {D'Ambrosio}}, \bibinfo {author} {\bibfnamefont {F.}~\bibnamefont {Vischi}},
  \bibinfo {author} {\bibfnamefont {F.~S.}\ \bibnamefont {Bergeret}}, \bibinfo
  {author} {\bibfnamefont {Y.~V.}\ \bibnamefont {Nazarov}}, and\ \bibinfo
  {author} {\bibfnamefont {F.}~\bibnamefont {Giazotto}},\ }\bibfield  {title}
  {\bibinfo {title} {{The $\omega$-SQUIPT: phase-engineering of Josephson
  topological materials}},\ }\href {https://doi.org/10.1038/nnano.2016.157}
  {\bibfield  {journal} {\bibinfo  {journal} {Nat. Nanotechnol.}\ }\textbf
  {\bibinfo {volume} {11}},\ \bibinfo {pages} {1055} (\bibinfo {year}
  {2016})}\BibitemShut {NoStop}%
\bibitem [{\citenamefont {Deb}\ \emph {et~al.}(2018)\citenamefont {Deb},
  \citenamefont {Sengupta},\ and\ \citenamefont {Sen}}]{Deb2018}%
  \BibitemOpen
  \bibfield  {author} {\bibinfo {author} {\bibfnamefont {O.}~\bibnamefont
  {Deb}}, \bibinfo {author} {\bibfnamefont {K.}~\bibnamefont {Sengupta}}, and\
  \bibinfo {author} {\bibfnamefont {D.}~\bibnamefont {Sen}},\ }\bibfield
  {title} {\bibinfo {title} {{Josephson junctions of multiple superconducting
  wires}},\ }\href {https://doi.org/10.1103/PhysRevB.97.174518} {\bibfield
  {journal} {\bibinfo  {journal} {Phys. Rev. B}\ }\textbf {\bibinfo {volume}
  {97}},\ \bibinfo {pages} {174518} (\bibinfo {year} {2018})}\BibitemShut
  {NoStop}%
\bibitem [{\citenamefont {Nowak}\ \emph {et~al.}(2019)\citenamefont {Nowak},
  \citenamefont {Wimmer},\ and\ \citenamefont {Akhmerov}}]{Nowak2019}%
  \BibitemOpen
  \bibfield  {author} {\bibinfo {author} {\bibfnamefont {M.~P.}\ \bibnamefont
  {Nowak}}, \bibinfo {author} {\bibfnamefont {M.}~\bibnamefont {Wimmer}}, and\
  \bibinfo {author} {\bibfnamefont {A.~R.}\ \bibnamefont {Akhmerov}},\
  }\bibfield  {title} {\bibinfo {title} {{Supercurrent carried by
  nonequilibrium quasiparticles in a multiterminal Josephson junction}},\
  }\href {https://doi.org/10.1103/PhysRevB.99.075416} {\bibfield  {journal}
  {\bibinfo  {journal} {Phys. Rev. B}\ }\textbf {\bibinfo {volume} {99}},\
  \bibinfo {pages} {075416} (\bibinfo {year} {2019})}\BibitemShut {NoStop}%
\bibitem [{\citenamefont {Pankratova}\ \emph {et~al.}()\citenamefont
  {Pankratova}, \citenamefont {Lee}, \citenamefont {Kuzmin}, \citenamefont
  {Vavilov}, \citenamefont {Wickramasinghe}, \citenamefont {Mayer},
  \citenamefont {Yuan}, \citenamefont {Shabani},\ and\ \citenamefont
  {Manucharyan}}]{Pankratova2018}%
  \BibitemOpen
  \bibfield  {author} {\bibinfo {author} {\bibfnamefont {N.}~\bibnamefont
  {Pankratova}}, \bibinfo {author} {\bibfnamefont {H.}~\bibnamefont {Lee}},
  \bibinfo {author} {\bibfnamefont {R.}~\bibnamefont {Kuzmin}}, \bibinfo
  {author} {\bibfnamefont {M.}~\bibnamefont {Vavilov}}, \bibinfo {author}
  {\bibfnamefont {K.}~\bibnamefont {Wickramasinghe}}, \bibinfo {author}
  {\bibfnamefont {W.}~\bibnamefont {Mayer}}, \bibinfo {author} {\bibfnamefont
  {J.}~\bibnamefont {Yuan}}, \bibinfo {author} {\bibfnamefont {J.}~\bibnamefont
  {Shabani}}, and\ \bibinfo {author} {\bibfnamefont {V.~E.}\ \bibnamefont
  {Manucharyan}},\ }\bibfield  {title} {\bibinfo {title} {{The multi-terminal
  Josephson effect}},\ }\href {http://arxiv.org/abs/1812.06017} {\ }\Eprint
  {https://arxiv.org/abs/1812.06017} {arXiv:1812.06017} \BibitemShut {NoStop}%
\bibitem [{\citenamefont {Qi}\ \emph {et~al.}(2018)\citenamefont {Qi},
  \citenamefont {Xie}, \citenamefont {Shabani}, \citenamefont {Manucharyan},
  \citenamefont {Levchenko},\ and\ \citenamefont {Vavilov}}]{Qi2018}%
  \BibitemOpen
  \bibfield  {author} {\bibinfo {author} {\bibfnamefont {Z.}~\bibnamefont
  {Qi}}, \bibinfo {author} {\bibfnamefont {H.~Y.}\ \bibnamefont {Xie}},
  \bibinfo {author} {\bibfnamefont {J.}~\bibnamefont {Shabani}}, \bibinfo
  {author} {\bibfnamefont {V.~E.}\ \bibnamefont {Manucharyan}}, \bibinfo
  {author} {\bibfnamefont {A.}~\bibnamefont {Levchenko}}, and\ \bibinfo
  {author} {\bibfnamefont {M.~G.}\ \bibnamefont {Vavilov}},\ }\bibfield
  {title} {\bibinfo {title} {{Controlled-Z gate for transmon qubits coupled by
  semiconductor junctions}},\ }\href
  {https://doi.org/10.1103/PhysRevB.97.134518} {\bibfield  {journal} {\bibinfo
  {journal} {Phys. Rev. B}\ }\textbf {\bibinfo {volume} {97}},\ \bibinfo
  {pages} {134518} (\bibinfo {year} {2018})}\BibitemShut {NoStop}%
\bibitem [{\citenamefont {Casparis}\ \emph {et~al.}(2018)\citenamefont
  {Casparis}, \citenamefont {Connolly}, \citenamefont {Kjaergaard},
  \citenamefont {Pearson}, \citenamefont {Kringh{\o}j}, \citenamefont {Larsen},
  \citenamefont {Kuemmeth}, \citenamefont {Wang}, \citenamefont {Thomas},
  \citenamefont {Gronin}, \citenamefont {Gardner}, \citenamefont {Manfra},
  \citenamefont {Marcus},\ and\ \citenamefont {Petersson}}]{Casparis2018}%
  \BibitemOpen
  \bibfield  {author} {\bibinfo {author} {\bibfnamefont {L.}~\bibnamefont
  {Casparis}}, \bibinfo {author} {\bibfnamefont {M.~R.}\ \bibnamefont
  {Connolly}}, \bibinfo {author} {\bibfnamefont {M.}~\bibnamefont
  {Kjaergaard}}, \bibinfo {author} {\bibfnamefont {N.~J.}\ \bibnamefont
  {Pearson}}, \bibinfo {author} {\bibfnamefont {A.}~\bibnamefont
  {Kringh{\o}j}}, \bibinfo {author} {\bibfnamefont {T.~W.}\ \bibnamefont
  {Larsen}}, \bibinfo {author} {\bibfnamefont {F.}~\bibnamefont {Kuemmeth}},
  \bibinfo {author} {\bibfnamefont {T.}~\bibnamefont {Wang}}, \bibinfo {author}
  {\bibfnamefont {C.}~\bibnamefont {Thomas}}, \bibinfo {author} {\bibfnamefont
  {S.}~\bibnamefont {Gronin}}, \bibinfo {author} {\bibfnamefont {G.~C.}\
  \bibnamefont {Gardner}}, \bibinfo {author} {\bibfnamefont {M.~J.}\
  \bibnamefont {Manfra}}, \bibinfo {author} {\bibfnamefont {C.~M.}\
  \bibnamefont {Marcus}}, and\ \bibinfo {author} {\bibfnamefont {K.~D.}\
  \bibnamefont {Petersson}},\ }\bibfield  {title} {\bibinfo {title}
  {{Superconducting gatemon qubit based on a proximitized two-dimensional
  electron gas}},\ }\href {https://doi.org/10.1038/s41565-018-0207-y}
  {\bibfield  {journal} {\bibinfo  {journal} {Nat. Nanotechnol.}\ }\textbf
  {\bibinfo {volume} {13}},\ \bibinfo {pages} {915} (\bibinfo {year}
  {2018})}\BibitemShut {NoStop}%
\bibitem [{\citenamefont {Riwar}\ \emph {et~al.}(2016)\citenamefont {Riwar},
  \citenamefont {Houzet}, \citenamefont {Meyer},\ and\ \citenamefont
  {Nazarov}}]{Riwar2016}%
  \BibitemOpen
  \bibfield  {author} {\bibinfo {author} {\bibfnamefont {R.~P.}\ \bibnamefont
  {Riwar}}, \bibinfo {author} {\bibfnamefont {M.}~\bibnamefont {Houzet}},
  \bibinfo {author} {\bibfnamefont {J.~S.}\ \bibnamefont {Meyer}}, and\
  \bibinfo {author} {\bibfnamefont {Y.~V.}\ \bibnamefont {Nazarov}},\
  }\bibfield  {title} {\bibinfo {title} {{Multi-terminal Josephson junctions as
  topological matter}},\ }\href {https://doi.org/10.1038/ncomms11167}
  {\bibfield  {journal} {\bibinfo  {journal} {Nat. Commun.}\ }\textbf {\bibinfo
  {volume} {7}},\ \bibinfo {pages} {11167} (\bibinfo {year}
  {2016})}\BibitemShut {NoStop}%
\bibitem [{\citenamefont {{J. S. Meyer }}\ and\ \citenamefont
  {Houzet}(2017)}]{Meyer2017}%
  \BibitemOpen
  \bibfield  {author} {\bibinfo {author} {\bibnamefont {{J. S. Meyer }}}and\
  \bibinfo {author} {\bibfnamefont {M.}~\bibnamefont {Houzet}},\ }\bibfield
  {title} {\bibinfo {title} {{Nontrivial Chern Numbers in Three-Terminal
  Josephson Junctions}},\ }\href
  {https://doi.org/10.1103/PhysRevLett.119.136807} {\bibfield  {journal}
  {\bibinfo  {journal} {Phys. Rev. Lett.}\ }\textbf {\bibinfo {volume} {119}},\
  \bibinfo {pages} {136807} (\bibinfo {year} {2017})}\BibitemShut {NoStop}%
\bibitem [{\citenamefont {Xie}\ \emph {et~al.}(2017)\citenamefont {Xie},
  \citenamefont {Vavilov},\ and\ \citenamefont {Levchenko}}]{Xie2017}%
  \BibitemOpen
  \bibfield  {author} {\bibinfo {author} {\bibfnamefont {H.~Y.}\ \bibnamefont
  {Xie}}, \bibinfo {author} {\bibfnamefont {M.~G.}\ \bibnamefont {Vavilov}},
  and\ \bibinfo {author} {\bibfnamefont {A.}~\bibnamefont {Levchenko}},\
  }\bibfield  {title} {\bibinfo {title} {{Topological Andreev bands in
  three-terminal Josephson junctions}},\ }\href
  {https://doi.org/10.1103/PhysRevB.96.161406} {\bibfield  {journal} {\bibinfo
  {journal} {Phys. Rev. B}\ }\textbf {\bibinfo {volume} {96}},\ \bibinfo
  {pages} {161406(R)} (\bibinfo {year} {2017})}\BibitemShut {NoStop}%
\bibitem [{\citenamefont {Xie}\ \emph {et~al.}(2018)\citenamefont {Xie},
  \citenamefont {Vavilov},\ and\ \citenamefont {Levchenko}}]{Xie2018}%
  \BibitemOpen
  \bibfield  {author} {\bibinfo {author} {\bibfnamefont {H.~Y.}\ \bibnamefont
  {Xie}}, \bibinfo {author} {\bibfnamefont {M.~G.}\ \bibnamefont {Vavilov}},
  and\ \bibinfo {author} {\bibfnamefont {A.}~\bibnamefont {Levchenko}},\
  }\bibfield  {title} {\bibinfo {title} {{Weyl nodes in Andreev spectra of
  multiterminal Josephson junctions: Chern numbers, conductances, and
  supercurrents}},\ }\href {https://doi.org/10.1103/PhysRevB.97.035443}
  {\bibfield  {journal} {\bibinfo  {journal} {Phys. Rev. B}\ }\textbf {\bibinfo
  {volume} {97}},\ \bibinfo {pages} {035443} (\bibinfo {year}
  {2018})}\BibitemShut {NoStop}%
\bibitem [{\citenamefont {Lee}\ \emph {et~al.}(2019{\natexlab{b}})\citenamefont
  {Lee}, \citenamefont {Shojaei}, \citenamefont {Pendharkar}, \citenamefont
  {Feldman}, \citenamefont {Mukherjee},\ and\ \citenamefont
  {Palmstr{\o}m}}]{Lee2019_2}%
  \BibitemOpen
  \bibfield  {author} {\bibinfo {author} {\bibfnamefont {J.~S.}\ \bibnamefont
  {Lee}}, \bibinfo {author} {\bibfnamefont {B.}~\bibnamefont {Shojaei}},
  \bibinfo {author} {\bibfnamefont {M.}~\bibnamefont {Pendharkar}}, \bibinfo
  {author} {\bibfnamefont {M.}~\bibnamefont {Feldman}}, \bibinfo {author}
  {\bibfnamefont {K.}~\bibnamefont {Mukherjee}}, and\ \bibinfo {author}
  {\bibfnamefont {C.~J.}\ \bibnamefont {Palmstr{\o}m}},\ }\bibfield  {title}
  {\bibinfo {title} {{Contribution of top barrier materials to high mobility in
  near-surface InAs quantum wells grown on GaSb(001)}},\ }\href
  {https://doi.org/10.1103/PhysRevMaterials.3.014603} {\bibfield  {journal}
  {\bibinfo  {journal} {Phys. Rev. Mater.}\ }\textbf {\bibinfo {volume} {3}},\
  \bibinfo {pages} {014603} (\bibinfo {year} {2019}{\natexlab{b}})}\BibitemShut
  {NoStop}%
\bibitem [{\citenamefont {McCumber}(1968)}]{McCumber1968}%
  \BibitemOpen
  \bibfield  {author} {\bibinfo {author} {\bibfnamefont {D.~E.}\ \bibnamefont
  {McCumber}},\ }\bibfield  {title} {\bibinfo {title} {{Effect of ac impedance
  on dc voltage-current characteristics of superconductor weak-link
  junctions}},\ }\href {https://doi.org/10.1063/1.1656743} {\bibfield
  {journal} {\bibinfo  {journal} {J. Appl. Phys.}\ }\textbf {\bibinfo {volume}
  {39}},\ \bibinfo {pages} {3113} (\bibinfo {year} {1968})}\BibitemShut
  {NoStop}%
\bibitem [{\citenamefont {Courtois}\ \emph {et~al.}(2008)\citenamefont
  {Courtois}, \citenamefont {Meschke}, \citenamefont {Peltonen},\ and\
  \citenamefont {Pekola}}]{Courtois2008}%
  \BibitemOpen
  \bibfield  {author} {\bibinfo {author} {\bibfnamefont {H.}~\bibnamefont
  {Courtois}}, \bibinfo {author} {\bibfnamefont {M.}~\bibnamefont {Meschke}},
  \bibinfo {author} {\bibfnamefont {J.~T.}\ \bibnamefont {Peltonen}}, and\
  \bibinfo {author} {\bibfnamefont {J.~P.}\ \bibnamefont {Pekola}},\ }\bibfield
   {title} {\bibinfo {title} {{Origin of hysteresis in a proximity josephson
  junction}},\ }\href {https://doi.org/10.1103/PhysRevLett.101.067002}
  {\bibfield  {journal} {\bibinfo  {journal} {Phys. Rev. Lett.}\ }\textbf
  {\bibinfo {volume} {101}},\ \bibinfo {pages} {067002} (\bibinfo {year}
  {2008})}\BibitemShut {NoStop}%
\bibitem [{\citenamefont {{De Cecco}}\ \emph {et~al.}(2016)\citenamefont {{De
  Cecco}}, \citenamefont {{Le Calvez}}, \citenamefont {Sac{\'{e}}p{\'{e}}},
  \citenamefont {Winkelmann},\ and\ \citenamefont {Courtois}}]{DeCecco2016}%
  \BibitemOpen
  \bibfield  {author} {\bibinfo {author} {\bibfnamefont {A.}~\bibnamefont {{De
  Cecco}}}, \bibinfo {author} {\bibfnamefont {K.}~\bibnamefont {{Le Calvez}}},
  \bibinfo {author} {\bibfnamefont {B.}~\bibnamefont {Sac{\'{e}}p{\'{e}}}},
  \bibinfo {author} {\bibfnamefont {C.~B.}\ \bibnamefont {Winkelmann}}, and\
  \bibinfo {author} {\bibfnamefont {H.}~\bibnamefont {Courtois}},\ }\bibfield
  {title} {\bibinfo {title} {{Interplay between electron overheating and ac
  Josephson effect}},\ }\href {https://doi.org/10.1103/PhysRevB.93.180505}
  {\bibfield  {journal} {\bibinfo  {journal} {Phys. Rev. B}\ }\textbf {\bibinfo
  {volume} {93}},\ \bibinfo {pages} {180505(R)} (\bibinfo {year}
  {2016})}\BibitemShut {NoStop}%
\bibitem [{\citenamefont {Tinkham}(1963)}]{Tinkham1963}%
  \BibitemOpen
  \bibfield  {author} {\bibinfo {author} {\bibfnamefont {M.}~\bibnamefont
  {Tinkham}},\ }\bibfield  {title} {\bibinfo {title} {{Effect of Fluxoid
  Quantization on Transitions of Superconducting Films}},\ }\href
  {https://doi.org/10.1103/PhysRev.129.2413} {\bibfield  {journal} {\bibinfo
  {journal} {Phys, Rev,}\ }\textbf {\bibinfo {volume} {129}},\ \bibinfo {pages}
  {2413} (\bibinfo {year} {1963})}\BibitemShut {NoStop}%
\bibitem [{\citenamefont {Maki}(1965)}]{Maki1965}%
  \BibitemOpen
  \bibfield  {author} {\bibinfo {author} {\bibfnamefont {K.}~\bibnamefont
  {Maki}},\ }\bibfield  {title} {\bibinfo {title} {{Fluxoid structure in
  superconducting films}},\ }\href
  {https://doi.org/10.1016/0003-4916(65)90153-3} {\bibfield  {journal}
  {\bibinfo  {journal} {Ann. Phys. (N.Y.)}\ }\textbf {\bibinfo {volume} {34}},\
  \bibinfo {pages} {3} (\bibinfo {year} {1965})}\BibitemShut {NoStop}%
\bibitem [{\citenamefont {Polonsky}\ \emph {et~al.}(1991)\citenamefont
  {Polonsky}, \citenamefont {Semenov},\ and\ \citenamefont
  {Shevchenko}}]{Polonsky1991}%
  \BibitemOpen
  \bibfield  {author} {\bibinfo {author} {\bibfnamefont {S.~V.}\ \bibnamefont
  {Polonsky}}, \bibinfo {author} {\bibfnamefont {V.~K.}\ \bibnamefont
  {Semenov}}, and\ \bibinfo {author} {\bibfnamefont {P.~N.}\ \bibnamefont
  {Shevchenko}},\ }\bibfield  {title} {\bibinfo {title} {{PSCAN: personal
  superconductor circuit analyser}},\ }\href
  {https://doi.org/10.1088/0953-2048/4/11/031} {\bibfield  {journal} {\bibinfo
  {journal} {Supercond. Sci. and Technol.}\ }\textbf {\bibinfo {volume} {4}},\
  \bibinfo {pages} {667} (\bibinfo {year} {1991})}\BibitemShut {NoStop}%
\bibitem [{Sup()}]{Supp}%
  \BibitemOpen
  \href@noop {} {}\bibinfo {note} {See Supplemental Material at [URL will be
  inserted by publisher] for a table of RCSJ parameters and simulation
  details.}\BibitemShut {Stop}%
\bibitem [{\citenamefont {Fourie}(2018)}]{Fourie2018}%
  \BibitemOpen
  \bibfield  {author} {\bibinfo {author} {\bibfnamefont {C.~J.}\ \bibnamefont
  {Fourie}},\ }\bibfield  {title} {\bibinfo {title} {{Digital Superconducting
  Electronics Design Tools-Status and Roadmap}},\ }\href
  {https://doi.org/10.1109/TASC.2018.2797253} {\bibfield  {journal} {\bibinfo
  {journal} {IEEE Trans. Appl. Supercond.}\ }\textbf {\bibinfo {volume} {28}},\
  \bibinfo {pages} {1300412} (\bibinfo {year} {2018})}\BibitemShut {NoStop}%
\bibitem [{\citenamefont {Altshuler}\ and\ \citenamefont
  {Garcı́a}(2003)}]{Altshuler2003}%
  \BibitemOpen
  \bibfield  {author} {\bibinfo {author} {\bibfnamefont {E.}~\bibnamefont
  {Altshuler}}and\ \bibinfo {author} {\bibfnamefont {R.}~\bibnamefont
  {Garcı́a}},\ }\bibfield  {title} {\bibinfo {title} {{Josephson junctions in
  a magnetic field: Insights from coupled pendula}},\ }\href
  {https://doi.org/10.1119/1.1533052} {\bibfield  {journal} {\bibinfo
  {journal} {Am. J. Phys.}\ }\textbf {\bibinfo {volume} {71}},\ \bibinfo
  {pages} {4} (\bibinfo {year} {2003})}\BibitemShut {NoStop}%
\bibitem [{\citenamefont {{L.W. Nagel }}\ and\ \citenamefont
  {Pederson}(1973)}]{Nagel1973}%
  \BibitemOpen
  \bibfield  {author} {\bibinfo {author} {\bibnamefont {{L.W. Nagel }}}and\
  \bibinfo {author} {\bibfnamefont {D.~O.}\ \bibnamefont {Pederson}},\ }\href
  {http://www2.eecs.berkeley.edu/Pubs/TechRpts/1973/22871.html} {\emph
  {\bibinfo {title} {{SPICE (Simulation Program with Integrated Circuit
  Emphasis)}}}},\ \bibinfo {type} {Tech. Rep.}\ \bibinfo {number} {UCB/ERL
  M382}\ (\bibinfo  {institution} {EECS Department, University of California,
  Berkeley},\ \bibinfo {year} {1973})\BibitemShut {NoStop}%
\end{thebibliography}%

\end{document}